\pdfoutput=1
\documentclass[nofootinbib, twocolumn, prd, preprintnumbers]{revtex4-1}
\usepackage{amsmath,amssymb,graphicx,booktabs,bm,psfrag,color,slashed,subfigure,mhchem,array}

\usepackage[dvipsnames]{xcolor}

\usepackage{etoolbox} 
    
\usepackage{siunitx} 
\DeclareSIUnit{\year}{y}

\begin{document}
\preprint{IPPP/20/99}

\vspace{1mm}
\title{\boldmath
Relic neutrinos at accelerator experiments}

\author{Martin Bauer}  
\author{Jack D. Shergold}

\affiliation{Institute for Particle Physics Phenomenology, Department of Physics, Durham University, Durham,
DH1 3LE, UK}

\begin{abstract}
We present a new technique for observing low energy neutrinos with the aim of detecting the cosmic neutrino background using ion storage rings. Utilising high energy targets exploits the quadratic increase in the neutrino capture cross section with beam energy, and with sufficient beam energy, enables neutrino capture through inverse-beta decay processes from a stable initial state. We also show that there exist ion systems admitting resonant neutrino capture, capable of achieving larger capture cross sections at lower beam energies than their non-resonant counterparts. We calculate the neutrino capture rate and the optimal experimental runtime for a range of different resonant processes and target ions and we demonstrate that the resonant capture experiment can be performed with beam energies as low as $\mathcal{O}(10\,\mathrm{TeV})$ per target nucleon. Unfortunately, none of the ion systems discussed here can provide sufficient statistics to discover the cosmic neutrino background with current technology. We address the challenges associated with realising this experiment in the future, taking into account the uncertainty in the beam and neutrino momentum distributions, synchrotron radiation, as well as the beam stability.
\end{abstract}

\maketitle
\section{Introduction}

\label{sec:intro}
\begin{figure}[t]
    \centering
 \includegraphics[width = 1.0\linewidth]{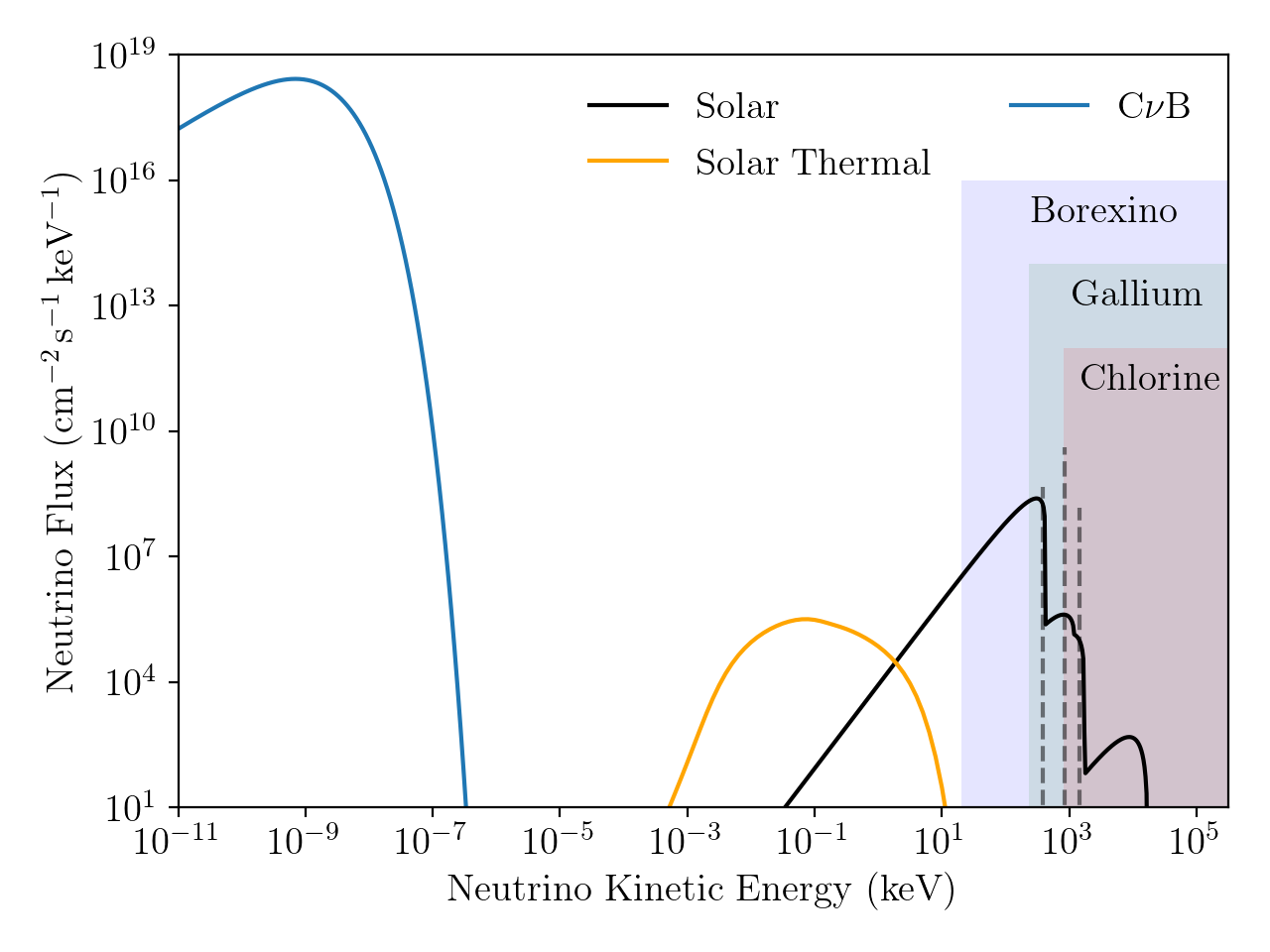}
 \caption{Lab frame fluxes of low-energy electron neutrinos with mass $m_\nu = 0.1\,\mathrm{eV}$, alongside the thresholds for Borexino and radiochemical detection experiments. The dashed lines have units \SI{}{\per\centi\metre\squared\per\second} and correspond to (from left to right): $^7$Be decay to $^7$Li$^*$ and $^7$Li, and $pep$ neutrinos. A full discussion of the solar fluxes is given in appendix~\ref{App:solarneutrinos}, and a comprehensive review of neutrino fluxes at all energy scales can be found in \cite{Vitagliano:2019yzm}.}
    \label{fig:fluxes}
\end{figure}

Neutrinos travel through the cosmos with minimal interactions with their environment, which makes them clean messengers of solar, astrophysical and cosmic phenomena. The weak interaction responsible for this unique property is also the leading cause of difficulties when attempting to directly detect neutrinos in a lab. Low-energy neutrinos are particularly challenging in this respect, making a future detection of the cosmic neutrino background (C$\nu$B) the holy grail of neutrino physics. Established detection techniques typically have a threshold of $\sim 300$ keV~\cite{Formaggio:2013kya}, as illustrated in Figure~\ref{fig:fluxes} and are therefore incapable of detecting neutrinos from the C$\nu$B, which have mean kinetic energy $E_{k,\nu}^0=1.8\times10^{-9}\,\mathrm{keV}$. 

The most discussed proposal to detect the C$\nu$B was originally suggested by Weinberg~\cite{Weinberg:1962zza}, which looks for an excess of events beyond the tritium beta decay electron endpoint energy, arising from neutrino capture process ${^3\text{H} + \nu_e \to {^3\text{He}^+} + e^-}$. Currently, KATRIN~\cite{Osipowicz:2001sq} is the most advanced experiment implementing this technique and is expected to set the best constraint on C$\nu$B overdensities, whilst future experiments such as HOLMES~\cite{DeGerone:2020jpq}, Project-8~\cite{Project8:2017nal} and ECHo~\cite{Gastaldo:2013wha} could improve upon these limits. In the future, a dedicated experiment such as PTOLEMY~\cite{Baracchini:2018wwj} could reach sensitivity to the C$\nu$B. Alternative proposals to detect the C$\nu$B include the observation of the $Z$-resonance absorption dip in the spectrum of ultra high-energy neutrinos~\cite{Fargion:1997ft,Eberle:2004ua}, utilising the neutrino wind using a torsion balance~\cite{Hagmann:1998nz,Shvartsman:1982sn}, laser interferometry experiments~\cite{Domcke:2017aqj}, the effect of Pauli blocking on electroweak de-excitation in atoms~\cite{Yoshimura:2014hfa} as well as the observation of coherent scattering of C$\nu$B neutrinos at accelerators~\cite{Ringwald:2004te}. 

We explore the sensitivity of an ion storage ring exploiting resonant, neutrino-induced beta decays and electron captures to detect cosmic neutrinos. Such an experiment could perform various measurements. The same exothermic process on which PTOLEMY is based can be observed in accelerated tritium ions~\cite{Ringwald:2004te} 
\begin{equation} \label{eq:forwardProcess}
^3\text{H}^+ + \nu_e \to {^3\text{He}^{++}} + e^-\,.
\end{equation}
At high energies, the neutrino capture cross section scales quadratically with the beam energy, allowing for significantly enhanced event rates. However, due to the natural beta decays of tritium, the signal can be overwhelmed by background. To overcome this, we may instead consider inverse beta decay induced by neutrino capture, \emph{e.g.} 
\begin{align} \label{eq:inverseProcess}
^3\text{He}^{++}+ \bar \nu_e \to {^3\text{H}^{+}} + e^+\,.
\end{align}
Because of the threshold of $Q_{\mathrm{He}}=18.59\,$keV in this example, this process can \emph{only} occur if an electron neutrino with sufficient energy is absorbed. For neutrinos of mass $m_\nu$ at rest\footnote{The mass of the electron neutrino is related to those of the mass eigenstates by $m_\nu = \sum_i |U_{ei}|^2\, m_{\nu_i}$.} and ions of mass $M$ accelerated to energy $E$ in the lab frame, this threshold corresponds to a neutrino energy in the rest-frame of the beam
\begin{align}\label{eq:Enu}
\widetilde{E}_{\nu} = \frac{m_\nu}{M}E \geq Q_{\mathrm{He}}\,.
\end{align}
Here and in the remainder of the paper we denote quantities in the beam frame with a tilde.  
For $^3$He,~\eqref{eq:Enu} translates into $E \geq 56 \,(\text{eV}/m_\nu) $ TeV, putting it in reach of a next-generation 100 TeV collider for $m_\nu \geq 0.56\,\mathrm{eV}$.
As the cross sections for both processes \eqref{eq:forwardProcess} and \eqref{eq:inverseProcess} scale as $G_F^2$, they are very small unless $\widetilde{E}_\nu$ significantly exceeds the electron mass. In addition the charge radius of the final state is different from the initial state, causing it to be ejected from the beam. We argue that both issues can be resolved by instead considering resonant $2\to 1$ processes, \emph{e.g.} 
\begin{align}{\label{eq:HeCap}}
{^3\text{He}^{++}+ \bar \nu_e + e^-(\text{bound}) \to {^3\text{H}^{+}} }\,,
\end{align}
and the cross section is maximised at the threshold. At resonance, the cross section for the process in \eqref{eq:HeCap} can exceed the at rest neutrino capture cross section by many orders of magnitude. For example, in the case of resonant capture of neutrinos produced in beta beams, the neutrino capture cross section for tritium exceeds the PTOLEMY cross section by seven orders of magnitude~\cite{Oldeman:2009wa}. This technique can be used to capture low-energy neutrinos using accelerated ions instead. 
The remainder of this paper is structured as follows.
We describe the sources of low-energy neutrinos in Section~\ref{sec:LEN} and explore non-resonant neutrino capture on accelerated ions in Section~\ref{sec:betabeam}. In Section~\ref{sec:resonant} we show how utilising resonant neutrino capture leads to larger cross sections compared to the non-resonant case and present novel 3-state systems with a clean, stable final state. In Section~\ref{sec:realworld} we calculate the beam conversion rates for a number of ion systems and address the challenges of realising a future experiment in Section~\ref{sec:beamLimitations}. We conclude in Section~\ref{sec:conclusions}.

\section{Low energy neutrinos}\label{sec:LEN}

The spectrum of low-energy neutrinos with an energy below the threshold of existing experiments is dominated by two sources, solar neutrinos and relic neutrinos of the C$\nu$B. While both can be targets of neutrino capture experiments we focus on the C$\nu$B.



Relic neutrinos decoupled from the Standard Model (SM) thermal bath at a temperature $T_\nu^\text{dec}\sim 1\,$MeV, and have since redshifted to $T_\nu^0\simeq 1.68\times 10^{-4}\,  $eV~\cite{Zeldovich:1981wf}. This
corresponds to a mean momentum of $p_\nu^0 = 0.529\,\text{meV}$, which we consider negligible compared with the neutrino mass~\cite{Long:2014zva}. The local, present day number density of relic neutrinos per degree of freedom can then be written as~\cite{Bernstein:1988ad}
\begin{align}
n_{\nu_i} = f_{c,i} \, n_0 \simeq  f_{c,i}\,56\,\text{cm}^{-3},
\end{align}
where $i=1,2,3$ and $f_{c,i}$ is a neutrino mass-dependent enhancement factor capturing local overdensities from gravitational clustering of non-relativistic neutrinos. Results from N-1-body simulations find enhancements of $f_{c,i}\simeq 1.1-1.2$ for neutrino masses $m_\nu < 0.1\,\mathrm{eV}$ up to 
a factor $f_{c,i}\simeq 100-200$ for $m_\nu = 1\,\mathrm{eV}$~\cite{Ringwald:2004np, deSalas:2017wtt, Mertsch:2019qjv}.\footnote{The exact value depends mostly on the density profile of the Dark Matter halo. Gravitational interactions between neutrinos and of neutrinos on Dark Matter are neglected in N-1-body simulations.} In very optimistic scenarios enhancements of up to $f_{c,i}\simeq 10^3-10^6$ are considered based on the local baryon overdensity~\cite{Lazauskas:2007da, Faessler:2014bqa}. 

Following the argument of~\cite{Long:2014zva}, the C$\nu$B consists exclusively of left helicity neutrino states and right helicity antineutrino states in the absence of significant gravitational interactions, whilst for strongly clustered neutrinos the density is shared equally between both helicity states. In the case that neutrinos are Majorona fermions, the right helicity states contribute to the neutrino flux. For the remainder of this paper we leave $f_{c,i}$, and consequently the left and right helicity relic neutrino densities $n_{\nu_{L,i}}$ and $n_{\nu_{R,i}}$ as free parameters\footnote{These satisfy $n_{\nu_{L,i}} + n_{\nu_{R,i}} = n_{\nu_i}$ for Dirac neutrinos, and $n_{\nu_{L,i}} + n_{\nu_{R,i}} = 2n_{\nu_i}$ for Majorana neutrinos.}, unless specified otherwise.
Finally, we note that in the rest frame of a highly relativistic ion beam the neutrino number density is further enhanced by the beam Lorentz factor $\gamma = E/M$ due to length contraction, however the total number of neutrino captures is a frame independent quantity owing to the competing time dilation effect. 

The exact distribution function of relic neutrinos today is unknown. However, in the absence of significant interactions since decoupling it is reasonable to suggest that it will be the same as at decoupling, redshifted. Assuming a Fermi-Dirac distribution at decoupling, this leads to a distribution per degree of freedom~\cite{Gnedin:1997vn}
\begin{equation} \label{eq:cnbFlux}
    \frac{dn_{\nu_i}}{dp_\nu} =\frac{2 n_{\nu_i}}{3\zeta(3)T_{\mathrm{eff}}^3} \frac{p_{\nu}^2}{  e^{\frac{p_{\nu}}{T_{\mathrm{eff}}}}+1},
\end{equation}
when normalised to $n_{\nu_i}$. Here $p_{\nu}$ is the magnitude of the neutrino three momentum and $T_{\mathrm{eff}}$ is the effective neutrino temperature, which we take to be constant with $T_{\mathrm{eff}}\simeq T^0_\nu$. Additionally, we assume degenerate neutrino masses such that the neutrino momentum $p_\nu$ is the same for all mass and flavour eigenstates.

Summing over the mass eigenstates and multiplying with the lab frame neutrino velocity $v_\nu = p_{\nu}/E_{\nu}$, we arrive at the electron neutrino flux in the lab frame

\begin{equation}
    \frac{d\phi_{\nu_e}}{dp_{\nu}} = \frac{p_{\nu}}{E_{\nu}} \sum_{i=1,2,3} |U_{ei}|^2\frac{dn_{\nu_i}}{dp_{\nu}},
\end{equation}
where $U_{ei}$ are the relevant entries of the Pontecorvo$\,$–$\,$Maki$\,$–$\,$Nakagawa$\,$-$\,$Sakata (PMNS) neutrino mixing matrix. The lab frame neutrino flux has standard deviation $\Delta_\nu  = 0.334\,\mathrm{meV}$, which we treat as the uncertainty in the momentum of any neutrino in the lab frame. We note that the value of $\Delta_\nu$ is approximately constant with the neutrino mass in the quasi-degenerate regime $m_\nu \gtrsim 0.1\,\mathrm{meV}$.

\begin{figure}[t]
    \centering
    \includegraphics[width = \linewidth]{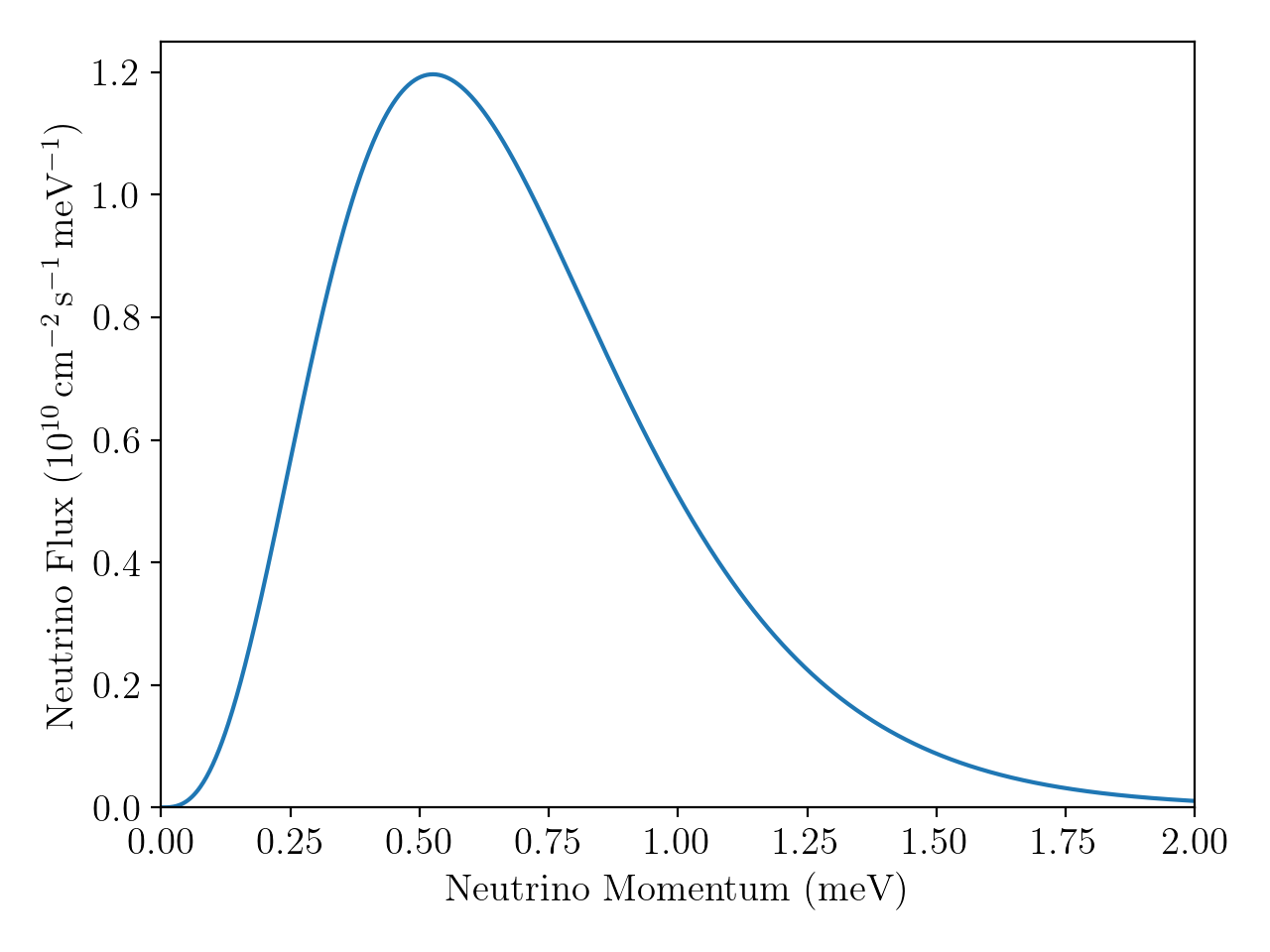}
    \caption{Lab frame electron neutrino flux from the C$\nu$B.}
    \label{fig:cnbPSpec}
\end{figure}

We propose to detect relic neutrinos by capturing them on a high-energy ion beam with velocity $v_b$. As a result, the factor $(v_b + v_\nu)n_{\nu_i}$ will enter into calculations of our event rates in the lab frame. Since $v_\nu \ll v_b \simeq 1$, we can define the effective total flux that the beam sees as $\phi_{\nu_e} \simeq v_b n_{\nu_i} \simeq n_{\nu_i}$. With this simplification, the lab and beam rest frame neutrino fluxes are related simply by $\widetilde \phi_{\nu_e} = \phi_{\nu_e}/\gamma$. Here the Lorentz factor $\gamma$ appears due to the length contraction of the direction parallel to the beam, increasing the neutrino number density in the beam rest frame. We plot the lab frame electron neutrino flux \eqref{eq:cnbFlux} in Figure~\ref{fig:cnbPSpec}.
%
%
%
%
In addition to relic neutrinos, we briefly discuss the possibility of observing solar neutrinos below the energy threshold of existing experiments such as Borexino ($20-200\,$keV~\cite{Alimonti:2008gc}). The distribution of low-energy solar neutrinos is depicted by the yellow contour in Figure~\ref{fig:fluxes}. While solar neutrinos are more energetic, the solar neutrino flux is smaller by several orders of magnitude compared to the expected flux of relic neutrinos. For our calculations of the solar neutrino flux we use the standard solar model given in~\cite{Asplund}, which we discuss further in appendix~\ref{App:solarneutrinos}. 

\section{Neutrino capture on a beam}\label{sec:betabeam}
The idea to capture relic neutrinos using tritium was first proposed by Weinberg, and the rate has been estimated based on our current best understanding of the C$\nu$B~\cite{Weinberg:1962zza, Akita:2020jbo}. In the following we explore the energy dependence of the neutrino capture cross section and show how it can be exploited for accelerated tritium or helium-3 to increase the event rate.\\
The cross section for neutrino capture on tritium for neutrinos with degenerate masses and helicity $\widetilde s_\nu$ is given by~\cite{Long:2014zva},
%
\begin{equation} \label{eq:tritiumCapture}
    \sigma\widetilde{v}_\nu  = (1-2\,\widetilde s_\nu \widetilde{v}_{\nu}){\sigma_0}(\widetilde{E}_e),
\end{equation}
where $\widetilde{v}_\nu$ is the neutrino velocity in the beam rest frame and the energy-dependent cross section reads
\begin{equation}\label{eq:sigbar}
    {\sigma_0}(\widetilde{E}_e) = G_F^2|V_{ud}|^2 F(2, \widetilde{E}_e) C(q^2) \frac{m_{^3\text{He}}}{m_{^3\text{H}}} \widetilde{E}_e \widetilde{p}_e\,.
\end{equation}
Here $\widetilde E_e$ and $\widetilde p_e$ are the energy and the magnitude of the final state electron three momentum, $V_{ud}$ is the Cabibbo$\,$–$\,$Kobayashi$\,$–$\,$Maskawa (CKM) matrix element and $C(q^2)$ contains details of nuclear structure.
We will use the value $C(q^2 \ll m_\pi) \simeq 0.874$ throughout.
The Fermi function encodes the enhancements due to electromagnetic effects between the final state nucleus and outgoing electron, and reads 
\begin{equation}\label{eq:fermi}
    F_-(Z,\widetilde{E}_e) = \frac{2(1+S)}{\Gamma(1+2S)^2}(2 \widetilde{p}_e \rho)^{2S-2} e^{-\pi \eta_e} |\Gamma(S - i\eta_e)|^2,
\end{equation}
where $\eta_e = Q_e Z\alpha/\widetilde{v}_e$ and $S = \sqrt{1-(\alpha Z)^2}$ are given in terms of the fine-structure constant $\alpha$, the charge of the final state electron $Q_e=-1$ and the atomic number of final state isotope $Z$, whilst $\rho \simeq 1.2 A^{\frac{1}{3}}\, \text{fm}$ is the nuclear radius. 
Summing over mass eigenstates and exploiting the unitarity of the PMNS matrix, we find that the total neutrino capture cross section on tritium at rest for non-relativistic neutrinos $\widetilde{\beta}_{\nu} = \widetilde{v}_\nu/c \ll 1$ is
\begin{equation}
        \sigma_{\text{NR}} = \sigma \widetilde{\beta}_{\nu} \Big|_{\widetilde{E}_e = E_0} 
        = 3.85\times 10^{-45} \,\text{cm}^2,
\end{equation}
where $E_0 = m_e + |Q_\mathrm{H}| + m_\nu$ is the energy carried by the final state electron, and $|Q_\mathrm{H}| = Q_\mathrm{He}$ is the energy released by the decay of tritium. This gives the total rate for the capture of Dirac neutrinos on $N_T^\text{rest}$ tritium targets at rest per year
\begin{equation} \label{eq:PtolemyRate}
    R_\text{rest}= n_\nu N_T^\text{rest} \sigma_{\text{NR}} \simeq 4f_{c,i}\,\mathrm{y}^{-1}
\end{equation}
where we have used $N_T^\mathrm{rest} = 2 \times 10^{25}$ in line with the PTOLEMY proposal~\cite{Baracchini:2018wwj}, corresponding to \SI{100}{\gram} of tritium. The capture rate for Majorana neutrinos is twice as large, as there will also be a flux of right-helicity neutrinos.\\
If instead tritium ions are accelerated on a beam, the larger centre of mass energy results in an enhanced neutrino capture cross section. Neglecting final state nuclear recoil, the electron energy and momentum in the tritium rest frame for accelerated tritium are given by 
\begin{align}
    \widetilde{E}_e &= m_e + |Q_\mathrm{H}| + \widetilde{E}_{\nu}, \\
    \widetilde{p}_e &= \sqrt{(\widetilde{E}_{\nu} + |Q_\mathrm{H}|)(\widetilde{E}_{\nu} + 2 m_e + |Q_\mathrm{H}|)} 
\end{align}
such that for $ \widetilde E_\nu > 2m_e$ and $\widetilde E_\nu > |Q_\mathrm{H}|$ the cross section increases quadratically with the neutrino energy $\sigma_0 \propto \widetilde E_\nu^2$. For an ion beam with energy $E$ and ion mass $M$, the energy of the neutrinos in the ion rest frame can then be written as $\widetilde{E}_{\nu}\simeq E m_\nu/M$, and the quadratic enhancement of the cross section becomes effective once the beam energy satisfies $E> 2m_e M/m_\nu$.
%
%
%
%
%
%
Exploiting this quadratic dependence is thus extremely challenging for the capture of cosmic neutrinos on tritium, as it requires a beam energy of 
\begin{equation}\label{eq:3Pev}
E\gtrsim\,  \SI{3}{\peta\electronvolt}\,\left[\frac{1\,\mathrm{eV}}{m_\nu}\right],
\end{equation}
for $M=m_{^3\rm{H}}\simeq 3$ GeV. The situation becomes more tenable when considering low energy solar neutrinos. For $E_\nu = $ $\SI{1}{\kilo\electronvolt}$ the quadratic enhancement begins at a beam energy of $E\gtrsim \SI{3}{\tera\electronvolt}$ for the same choice of target, well within reach of modern accelerators. However, unless the thermal solar neutrino flux significantly exceeds the predicted value shown in Figure~\ref{fig:fluxes}, the capture rate would be too low to observe at that energy scale.

\begin{figure}[t]
    \centering
     \includegraphics[width = \linewidth]{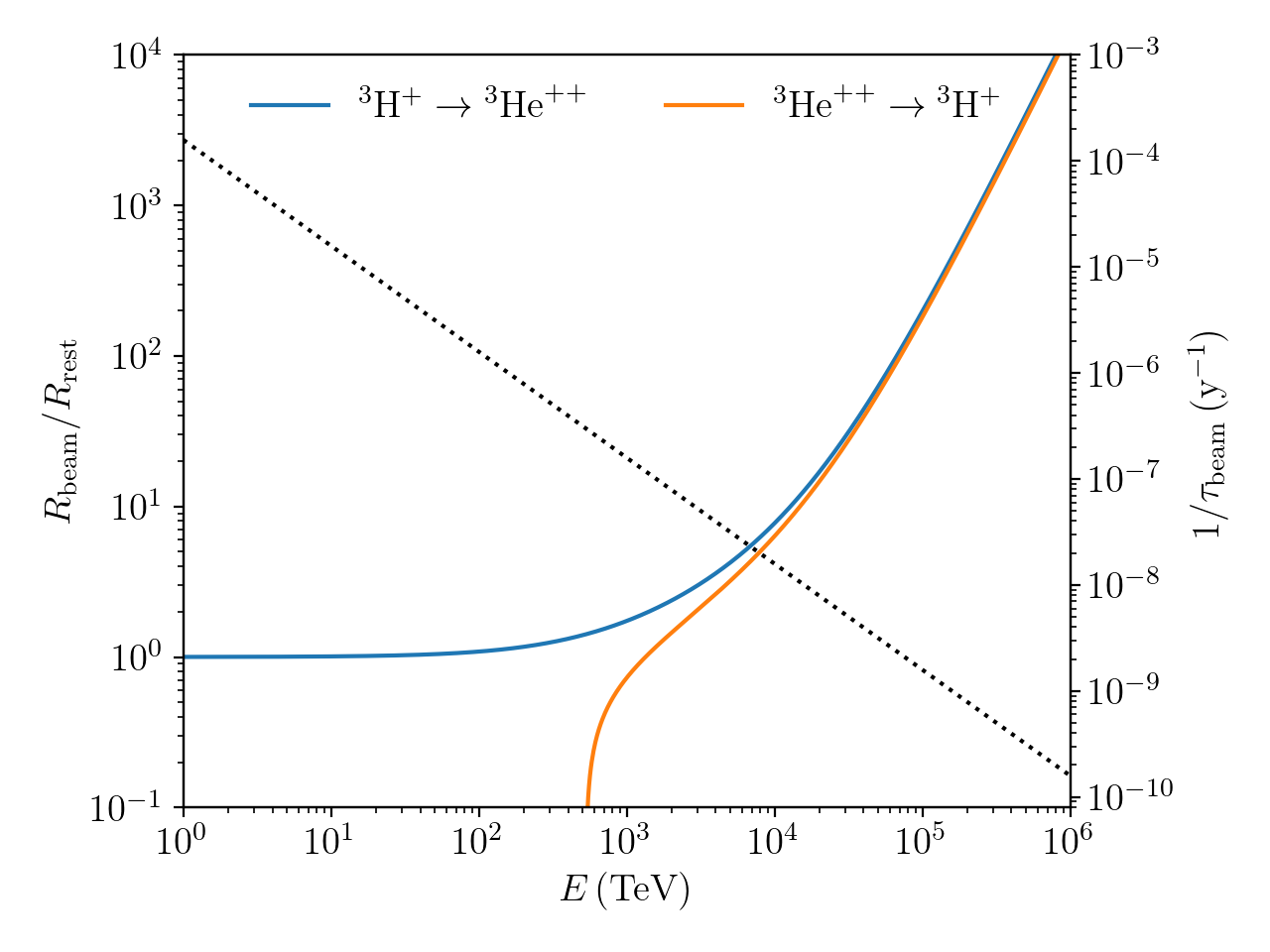}
    \caption{Event rate enhancement for a \textit{PTOLEMY-on-a-beam} type experiment (blue) as a function of the beam energy $E$ using $m_\nu=0.1$ eV, alongside the rate for the inverse process on a beam (orange), assuming $N_T^{\mathrm{beam}} = N_T^{\mathrm{rest}}$. Also shown by the dotted line is the decrease in the tritium decay rate as a function of the beam energy.}
    \label{fig:rateEnhancement}
\end{figure}

For an accelerated beam of tritium ions, the helicity dependent terms in \eqref{eq:tritiumCapture} can become important. Setting $\widetilde v_\nu\simeq v_\mathrm{rel} =1$, we find
\begin{equation}
    \sigma \widetilde{\beta}_\nu =
    \begin{cases}
    2 \,{\sigma_0}(\widetilde{E}_e), &\quad \widetilde s_\nu = -\frac{1}{2}\,,   \\[2pt]
    0, &\quad \widetilde s_\nu = +\frac{1}{2}\,,
    \end{cases}
\end{equation}
such that the neutrino capture cross section for right-helicity states vanishes identically, whilst the left-helicity capture cross section doubles. One would then naively expect the total neutrino capture rate to double with respect to the capture rate at rest~\eqref{eq:PtolemyRate}. However, due to the combined effects of Lorentz contraction and the velocity of the beam relative to the neutrinos, the left-helicity neutrino density for any neutrino flavor in the beam frame reads
\begin{equation}
    \widetilde{n}_{\nu_{L,i}} = \gamma \big(n_{\nu_{L,i}} + P_{-}(n_{\nu_{R,i}}-n_{\nu_{L,i}})\big),
\end{equation}
where $P_{-}$ is the probability that a relic neutrino will flip helicity when boosted into the beam frame
\begin{equation}
    P_- = \begin{cases}
    \displaystyle \frac{1}{2} - \frac{1}{\pi}\arcsin\left(\displaystyle \frac{\bar{v}_\nu}{\bar{v} \sin\bar{\theta}_\nu}\right), &\quad \bar{v} \sin \bar{\theta}_\nu > \bar{v}_\nu  \,, \\[2pt]
    0, &\quad \bar{v} \sin \bar{\theta}_\nu \leq \bar{v}_\nu\,,
    \end{cases}
\end{equation}
and $\bar{v}$ and $\bar{v}_\nu$ are the average beam and neutrino velocities in the lab frame, respectively, and $\bar{\theta}_\nu \in [0, \pi]$ is the average angle between the neutrinos and the axis perpendicular to the beam plane. We first consider Dirac neutrinos.
In the case of a relic neutrino background close to the prediction by standard cosmology $f_{c,i}\simeq 1$, they are randomly oriented in the lab frame,  $\bar{\theta}_\nu = \pi/2$,  and $n_{\nu_{R,i}} = 0$. Since the neutrinos are non-relativistic in the lab frame, $\bar{v}_\nu \ll \bar{v} \sin \bar{\theta}_\nu$, it follows that $P_{-} \simeq 1/2$ and $\widetilde{n}_{\nu_{L,i}} = \gamma n_{\nu_i}/2$. For large overdensities $f_{c,i}\gg 1$, the strong gravitational clustering leads to $n_{\nu_{L,i}} = n_{\nu_{R,i}} =n_{\nu_i}/2$, and we arrive at the same result. 
For Majorana neutrinos, $n_{\nu_{L,i}}= n_{\nu_{R,i}} =n_{\nu_i}$ independent of clustering, giving a left-helicity neutrino number density twice as large as that for Dirac neutrinos in the cases discussed above~\cite{Long:2014zva}. We consider Dirac neutrinos with $n_{\nu_{L,i}} =n_{\nu_i}/2$ for the remainder of the paper. 
We then arrive at the total capture rate for a number of target ions $N_T^\text{beam}$ on a beam, 
\begin{align}
R_\text{beam} =\frac{n_{\nu_i}}{2} N_T^\text{beam} \sigma_\text{R}(\widetilde{E}_e) , 
\end{align}
where the relativistic cross section can be written as
\begin{equation} \label{eq:sigr}
    \sigma_\text{R}(\widetilde{E}_e) = \frac{2\widetilde{E}_e \widetilde{p}_e}{E_0 \sqrt{E_0^2 - m_e^2}}\frac{F_-(2,\widetilde E_e)}{F_-(2,E_0)} \,\sigma_\text{NR}\,,
\end{equation}
and the ratio of Fermi functions is $\sim 1$.
In order for a \emph{PTOLEMY-on-a-beam} type experiment to improve the event rate of the experiment at rest one would therefore need 
\begin{align}\label{eq:RPOB}
\frac{R_\text{beam}}{R_\text{rest}} 
\simeq\bigg(\frac{E}{M}\bigg)^2\!\frac{m_\nu^2}{\sqrt{2|Q_\mathrm{H}|m_e^3}}\frac{N_T^\text{beam}}{2\times 10^{25}}>1\,,
\end{align}
where we assume~\eqref{eq:3Pev} holds and $N_T^\text{rest}= 100 N_A/3 $ with the Avogadro number $N_A\simeq$ \SI{6e23}{\per\mole} for the target at rest as proposed in~\cite{Betti:2019ouf}. In Figure~\ref{fig:rateEnhancement} we show the ratio~\eqref{eq:RPOB} for $N_T^\text{beam}=N_T^\text{rest} $ tritium ions as a function of the beam energy in blue, using $m_\nu=0.1$ eV. The quadratic enhancement becomes apparent at $E\gtrsim 30\,$PeV, which far exceeds that of any existing or proposed accelerators.

%
The reconstruction of signal events is additionally complicated by the fact that the final state $^3$He$^{++}$ is produced without interacting with background neutrinos by the natural beta decay of tritium $^3\text{H}^+\to \bar \nu_e + ^3\text{He}^{++}+e^-$ with lifetime $\tau_0 = 17.8\,\mathrm{y}$, producing an irreducible background to the signal process. This background is huge, even if one takes into account the relativistically increased lifetime of the beam ions $\tau_\text{beam}=\gamma \tau_0$. In Figure~\ref{fig:rateEnhancement} we show the decay rate $1/\tau_\text{beam}$ as a function of $E$ with the dotted line. 


This background can be avoided if the beam energy is large enough to enable the inverse process $^3\text{He}^{++}+~\bar\nu_e~\to~{^3\text{H}^{+}}~+~e^+$ which has a threshold $Q_\text{He}=18.59\,$ keV.  In this case the production of tritium ions in the final state provides a clean signal state, because
the final state can only be obtained by the anti-neutrino capture process, at the cost of a decaying signal. The cross section for the inverse process reads
%
\begin{equation}
        {\sigma_0^\text{inv}}(\widetilde{E}_e) = G_F^2|V_{ud}|^2 F_+(1, \widetilde{E}_e) C(q^2)  \frac{m_{^3\text{H}}}{m_{^3\text{He}}} \widetilde{E}_e \widetilde{p}_e,
    \label{eq:siginv}
\end{equation}
assuming the same form factors as the tritium process and defining $F_+(Z,\widetilde E_e)$ as \eqref{eq:fermi} for the positron with charge $Q_e=+1$. 

In the beam frame, the anti-neutrino energy required to overcome the threshold for the inverse process is $\widetilde E_{\nu} >Q_\text{He}$, and the electron energy and momentum are given by
\begin{align}
    \widetilde{E}_e &= m_e - Q_\text{He} + \widetilde{E}_{\nu}, \\
    \widetilde{p}_e &= \sqrt{(\widetilde{E}_{\nu} - Q_\text{He})(\widetilde{E}_{\nu} + 2 m_e - Q_\text{He})}\,.
\end{align}
For $m_{^3\text{H}}\simeq m_{^3\text{He}}$, and neglecting changes in the form factors, the enhancement for the relativistic cross section follows \eqref{eq:sigr} with the replacement $F_{-}(2,\widetilde{E}_e)\rightarrow F_{+}(1,\widetilde{E}_e)$. In the high-energy limit $\widetilde E_{\bar \nu}\gg 2 m_e$, the enhancement over PTOLEMY is then given by \eqref{eq:RPOB} to a very good approximation. This behaviour is reflected in Figure~\ref{fig:rateEnhancement} by the orange contour asymptotically tracing the blue contour for large $E$.

We chose the ${^3\mathrm{H}}-{^3\mathrm{He}}$ system to illustrate the neutrino capture processes in this section. In principle the cross section~\eqref{eq:sigbar} can be enhanced if one considers ions with smaller energy gaps, but the scaling with the initial state ion mass $M$ in \eqref{eq:RPOB} favours the ${^3\mathrm{H}}-{^3\mathrm{He}}$ system over any other choice. For an experiment with accelerated ions to realise a substantial enhancement over the neutrino capture rate at an experiment at rest like PTOLEMY requires beam energies orders of magnitude above what can be achieved by current accelerator technology. However, with sufficient energy, accelerated ion beams can capture cosmic neutrinos even if the capture process has a threshold. This leads to a clean final state consisting of ions that cannot be produced in an experiment at rest.
Then provided one can overcome the threshold $\widetilde E_\nu> Q$, the inverse process is always superior to the thresholdless process on a beam. However, as the tritium final state has a different charge radius to the initial state, the challenge of collecting the signal before it exits the beam remains. 

 \section{Resonant processes}\label{sec:resonant}

The main limitation identified in the last section for cosmic neutrino capture on a beam is the large energy required to achieve an enhancement over the same experiment at rest. 
It is not surprising that the neutrino capture cross section for accelerated tritium ions scales quadratically with the beam energy, because the cross section is suppressed by $G_F^2$. This $G_F^2$ suppression is a universal problem in proposals to detect the C$\nu$B~\cite{chiaki}. 
Coherent scattering effects cannot overcome this problem, as was established in a no-go theorem prohibiting cross sections scaling linearly in $G_F$~\cite{Cabibbo:1982bb, Langacker:1982ih}. The single exception is the Stodolsky effect which can only be measured for a net helicity asymmetry of the cosmic neutrino background~\cite{Stodolsky:1974aq}. An alternative method of avoiding $G_F^2$ suppression in the cross section is to use a resonance. 
Very energetic neutrinos can scatter off the cosmic neutrino background at the $Z$-resonance with a peak cross section of order $G_F$~\cite{Fargion:1997ft},
\begin{align}
\sigma_Z = \frac{12\pi}{M_Z^2} \mathrm{Br}(Z\to \bar{\nu}_\alpha \nu_\alpha) \propto G_F  \,.
\end{align}
This process has not been observed yet, but would lead to a dip in the high energy cosmic neutrino spectrum at $E_\nu \simeq 10^{22}$ eV for a neutrino mass $m_\nu=1$ eV, and the most energetic neutrino events seen today have an energy of $E_\nu \simeq 6\times 10^{15} $eV~\cite{IceCube:2021rpz}. Here we propose to use a beam of accelerated ions to capture neutrinos at nuclear $\beta$-resonances, resulting in a cross section that is independent of $G_F$. 
We consider two resonant processes: the resonant bound beta decay process (RB$\beta$)
\begin{align}
    {^{A}_{Z}P} + \nu_e &\rightarrow {^{A}_{Z+1}D} + e^-(\mathrm{bound}),
\end{align}
and the resonant electron capture process (REC)
\begin{align}
    {^{A}_{Z}P} + e^-(\mathrm{bound}) + \bar{\nu}_e &\rightarrow {^{A}_{Z-1}D},
\end{align}
for parent and daughter nuclei $P$ and $D$, respectively.    
The cross section for these processes involving relativistic beam frame neutrinos with energy $\widetilde{E}_\nu$ and resonance threshold $Q$ can be described by a Breit-Wigner function,
\begin{equation} \label{eq:rec}
        \sigma_\text{Res} = \frac{2\pi}{\widetilde{E}_{\nu}^2} \left(\frac{2 J_D + 1}{2J_P+1}\right)\bigg[\frac{\Gamma^2/4}{(\widetilde{E}_{\nu}-Q)^2 + \Gamma^2/4}\bigg] \mathcal{B}_{DP},
\end{equation}
where $\Gamma$ is the total width of the resonant state $D$, $J_P$ and $J_D$ denote the total angular momentum of the parent and daughter nuclei, and $\mathcal{B}_{DP}=\text{Br}(D\to P\,\nu_e)$ or $\mathcal{B}_{DP}=\text{Br}(D\to P\,\bar \nu_e e^-\text{(bound)} )$ are the branching ratios of the daughter $D$ to decay back to the parent state $P$. 
For a beam tuned to $\widetilde{E}_\nu \simeq \widetilde{p}_\nu = Q$ the cross section is resonantly enhanced, with a peak cross section
\begin{equation} \label{eq:resXsecNumeric}
    \sigma_\mathrm{peak} = 2.5\times 10^{-15}\, \frac{2 J_D + 1}{2J_P+1} \left[\frac{1\,\mathrm{keV}}{Q}\right]^2 \mathcal{B}_{DP}\,\mathrm{cm}^2,
\end{equation}
which is suppressed by $Q^2$ instead of $G_F$.
Expressed in terms of the neutral atom masses $M(Z,A)$, the thresholds for the resonant processes involving nuclei with degree of ionisation $I$ are given by

\begin{align}
    \begin{split}\label{eq:QREC}
        Q_{\text{REC}} = M(Z-1,A) &+ B_{I-1}(Z-1) - M(Z,A)\\ 
                            &- B_{I}(Z) + B_{\mathrm{EC}}(Z-1)\\
                            &+ E^*(Z-1,A) - E^*(Z,A),
    \end{split} \\
    \begin{split}\label{eq:QRBbeta}
        Q_{\text{RB}\beta} = M(Z+1,A) &+ B_{I+1}(Z+1) - M(Z,A)\\ 
                            &- B_{I}(Z) - B_{\mathrm{B}\beta}(Z+1)\\
                            &+ E^*(Z+1,A) - E^*(Z,A),
    \end{split}
\end{align}
where $B_I(Z)$ is the binding energy of the $I$ outermost electrons taken away from the neutral atom to produce the ionised state, while $B_{\mathrm{EC}}(Z)>0$ and $B_{\mathrm{B}\beta}(Z)>0$ are the binding energies of the missing and captured final state electrons, respectively. The remaining parameter, $E^*(Z)$, denotes the excitation energy of any nuclear isomers that may be chosen as the initial or resonant state, which can significantly affect the value of $Q$. In this section we will consider fully ionised RB$\beta$ initial states and REC initial states with a single $1s$ shell electron, such that we can neglect electron de-excitation contributions to $\mathcal{B}_{DP}$ in the resonance state.

The use of a state with at most $1s$ shell electrons is what allows for a cross section that is independent of $G_F$, as only weak decays are allowed such that $\mathcal{B}_{DP}$ is $\mathcal{O}(1)$. As a result, the cross section~\eqref{eq:resXsecNumeric} looks very promising, but the total width $\Gamma\propto G_F^2$ is narrow. 
The lab frame momentum distribution of relic neutrinos has a finite width, whilst the cross section to capture neutrinos away from resonance is suppressed by a factor $\Gamma^2/4E_{\nu}$. Therefore the resonant process will only capture the small fraction of neutrinos whose energy in the beam frame lies on a narrow interval about the resonance peak, a fact which is not helped by additional broadening due to a non-monochromatic ion beam.
For a very narrow resonance the neutrino capture rate $R$ per target ion $N_T$ can be written as
\begin{align}
    \frac{R}{N_T} &= \int\displaylimits_{Q}^{\infty}d\widetilde{E}_{\nu} \,\sigma_{\mathrm{Res}}(\widetilde{E}_{\nu}) \,\frac{d\phi_{\nu_e}}{d\widetilde{E}_{\nu}} \notag\\
    &\simeq \frac{\pi}{2} \sigma_{\mathrm{peak}} \Gamma \frac{d\phi_{\nu_e}}{d\widetilde{E}_{\nu}}\Big|_{\widetilde{E}_{\nu} = Q} &\quad 
   \notag \\
    &=\sqrt{\frac{\pi^3}{2}} \left(\frac{2 J_D + 1}{2J_P+1}\right) \frac{\Gamma}{Q^2} \frac{\mathcal{B}_{DP}}{\widetilde \Delta_E}\phi_{\nu_e}\,,
\end{align}
%
%
%
%
%
where in the last step we have assumed that the neutrino energy distribution in the beam frame is Gaussian with a standard deviation $\widetilde \Delta_E \gg \Gamma$. 

It is then instructive to attempt to estimate $\widetilde\Delta_E$. Given that the uncertainty in the lab frame neutrino momentum is $\Delta_\nu$ and the beam momentum is subject to fluctuations of scale $\Delta_b$, we find that 
\begin{equation}
   \widetilde \Delta_E = \sqrt{\bigg(\Delta_\nu\frac{\partial \widetilde{E}_{\nu}}{\partial p_\nu}\bigg)^2 + \bigg(\Delta_b \frac{\partial \widetilde{E}_{\nu}}{\partial p}\bigg)^2}.
  \end{equation}
For non-relativistic C$\nu$B neutrinos $p_\nu \ll m_\nu$ and $p \gg M$ this results in
\begin{equation}
   \widetilde \Delta_E \simeq \frac{1}{M}\sqrt{(E \Delta_\nu)^2 + (m_\nu \Delta_b)^2}= Q\sqrt{\delta_\nu^2 + \delta_b^2}\,,
\end{equation}
where we have defined the fractional uncertainties $\delta_b = \Delta_b/E$ and $\delta_\nu = \Delta_\nu/E_\nu$. This leads to the event rate for $\widetilde \Delta_E\gg \Gamma$
%
%
\begin{align}\label{eq:intRate}
    \frac{R}{N_T}&=\sqrt{\frac{\pi^3}{2}}\left(\frac{2 J_D + 1}{2J_P+1}\right) \frac{\Gamma}{Q^3} \frac{\mathcal{B}_{DP}}{\sqrt{\delta_\nu^2 + \delta_b^2}}\phi_{\nu_e}.
\end{align}
However, it is more convenient to define the \emph{quality factor} as the ratio of the rate for producing signal events over the rate of signal loss
\begin{align}\label{eq:Rtau}
   R_\tau&\equiv \frac{\gamma}{\Gamma}\frac{R}{N_T}\\
   &=1.7\times10^{-17}\, \frac{2 J_D + 1}{2J_P+1}\frac{\mathcal{B}_{DP}\, f_{c,i}}{\sqrt{\delta_\nu^2 + \delta_b^2}} \left[\frac{0.1\,\mathrm{eV}}{m_\nu}\right]\left[\frac{1\,\mathrm{keV}}{Q}\right]^2\!\!.\notag
\end{align}
In the following we will use this dimensionless quantity to measure how effective a given system of ions is at producing and retaining signal states on the beam.

\subsection{Resonant 2-state systems}\label{sec:2stateGS}
The most straightforward realisation of resonant neutrino capture on a beam is a 2-state system consisting of a parent ion $P$ and a resonant daughter state $D$ which decays back into the initial parent particle.
For resonant bound beta decay, the daughter ion will decay back through electron capture
\begin{align}\label{eq:process1}
\raisebox{-.4cm}{\includegraphics[scale=.48]{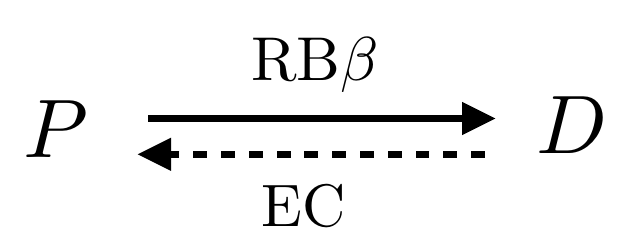}}.
\end{align}
Here and in the following processes contributing to the production of signal states are indicated by a solid arrow. Dashed arrows indicate processes that reduce the signal or the number of targets available to capture neutrinos, \emph{e.g.} by decays of the daughter ion back into the parent state.
In the case of resonant electron capture the daughter decays back into the parent ion by bound beta decay. A fraction of the daughter ions will also decay through continuous beta decay into a $P^+$ ion, which will be ejected from the beam eventually because of its charge-mass ratio. We can write this process as
\begin{align}\label{eq:process2}
\raisebox{-.3cm}{\includegraphics[scale=.48]{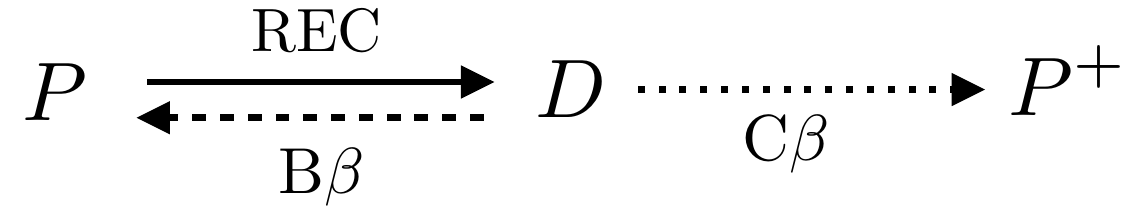}}\,.
\end{align}
In the case of resonant beta decay~\eqref{eq:process1}, the daughter ion can decay into different nuclear isomers of the parent ion. Only decays into the parent state isomer contribute to the branching ratio that appears in the resonant cross section~\eqref{eq:rec}. We parameterise the fraction of decays into the parent isomer over decays into any other isomer by $\chi \leq 1$. Here, $\chi$ accounts for the fact that some daughter ions might decay into excited parent states. In the case of electron capture the continuous beta decay of the daughter ion does not contribute to the resonant cross section either. We can then write for branching ratio
\begin{equation}
    \mathcal{B}_{DP} =  \begin{cases} \displaystyle
        \chi, &\quad \mathrm{RB}\beta, \\[2pt] \displaystyle{
        \frac{\chi}{1+K(Z,Q)}}, &\quad \mathrm{REC}.
    \end{cases}
\end{equation}
Here $K(Z,Q)$ denotes the ratio of the width for the daughter ion to bound beta decay into the $n=1$ shell with $n_f = 2$ vacant $s$-orbitals, over the width for continuous beta decay, defined in appendix~\ref{App:bbdecay}. 

%
For any given experiment we are interested in the number of signal states $N_D$ on the beam after a runtime $t$. In order to compare ions with vastly different lifetimes we introduce the dimensionless variable $x=t/\gamma\tau_D$, where $\tau_D$ is the lifetime of the daughter ion at rest and $\gamma$ the Lorentz factor of the beam. 
For an initial number of parent ions $N_0$ on the beam, the expected number of resonant daughter states $D$ on the beam is given by, 
\begin{align} \label{eq:xySignal0}
    N_D(x) = N_0 R_\tau (1 - e^{-x}) + \mathcal{O}(R_\tau^2)
\end{align}
where the quality factor is defined in~\eqref{eq:Rtau} and is considered to be $R_\tau\ll 1$. For completeness we give the full expressions for this and other beam population equations in appendix~\ref{App:beamstates}. Determining the rate of resonant neutrino capture in 2-state systems can therefore be achieved by solving \eqref{eq:xySignal0} for $R_\tau$ with a measured value of $N_D$. There are diminishing returns in running the experiment beyond $x>x_\mathrm{max}\simeq 1$, as $N_D(x \gtrsim x_\mathrm{max})\simeq N_0 R_\tau$. This corresponds to a dynamic equilibrium, where $D$ particles are produced by neutrino capture at same rate that they decay back into the initial state $P$. In the cases where $\mathcal{B}_{DP} < \chi$, no true equilibrium can ever be reached, as the total number of ions on the beam is depleted by decays $D\to P^+$. However, this effect is negligible for small $R_\tau$.

The large $x$ limit therefore represents the theoretical maximum number of resonances that can be on the beam at any one time, and suggests using short lifetime states such that $N_D$ reaches its maximum with a shorter experimental runtime. Unfortunately, this presents an experimental challenge, as short lifetime states will inevitably decay \emph{if} the beam needs to be slowed to make a measurement of $N_D$. 


\begin{figure}
    \centering
    \includegraphics[width=1.1\linewidth]{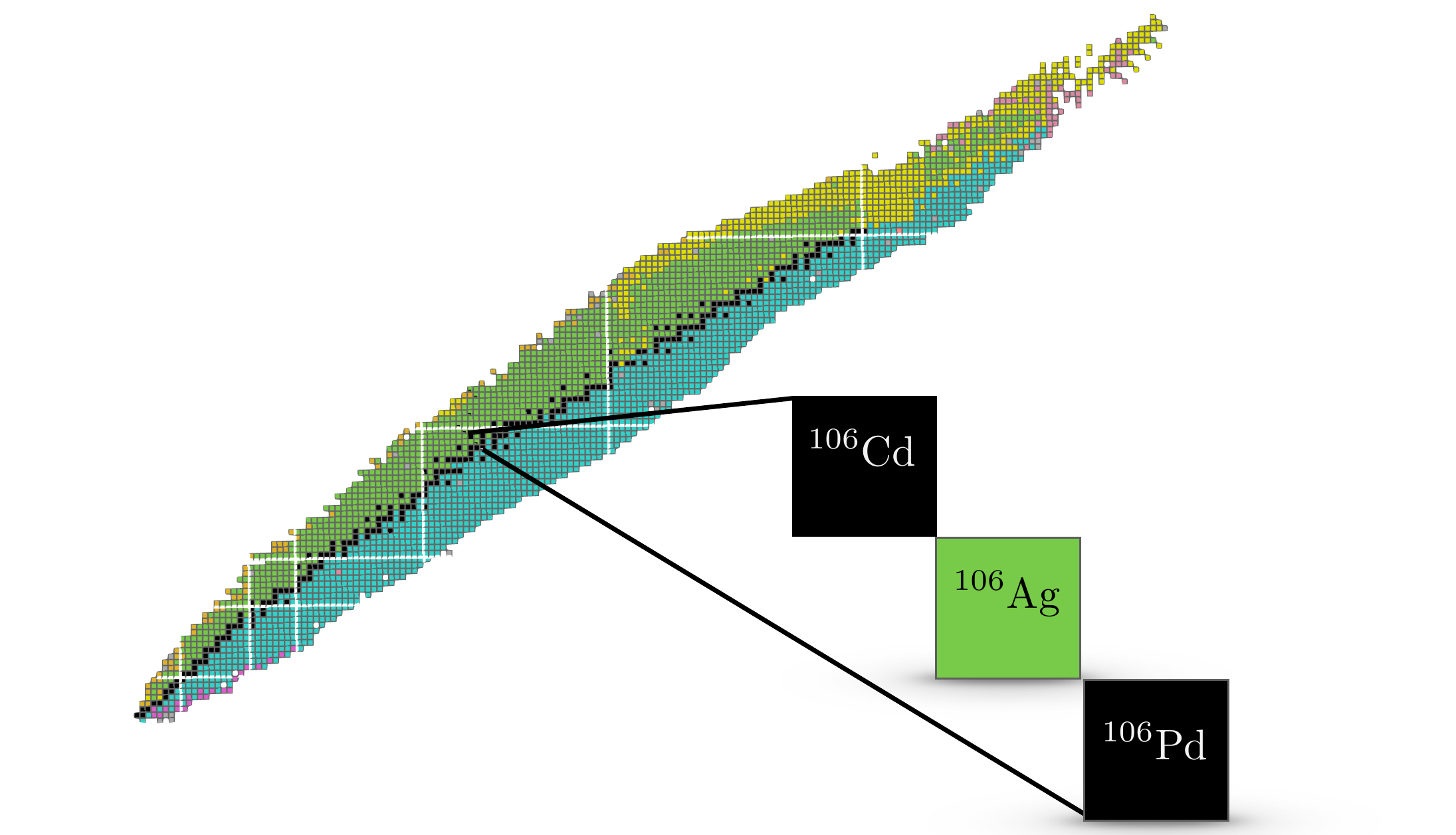}
    \caption{Isobar triplets with ``stable-unstable-stable" configurations are candidates for 3-state systems for resonant neutrino capture. Graphic taken in part from~\cite{iaeachart}}
    \label{fig:isotopes}
\end{figure}

\subsection{Resonant 3-state systems}

The conundrum of 2-state systems for resonant neutrino capture can be overcome using systems with a final state $F$ different from both the parent and daughter ions. We call such an arrangement a 3-state system. In 3-state systems the resonant daughter ion can retain a large decay branching ratio into the parent ion and therefore a large resonance can be achieved, but a third, \emph{stable} state can be produced through a different decay of the daughter ion. If this stable state has the same charge-mass ratio as both the $P$ and $D$ states, then it can remain on the beam indefinitely and can be detected after the run is finished. Candidate 3-state systems correspond to ``stable-unstable-stable" triplets of isobars that can be identified in the table of nuclides as illustrated in Figure~\ref{fig:isotopes}.

The analogous decay chain of the neutrino-induced resonant bound beta decay \eqref{eq:process1} for the 3-state system is  
\begin{align}\label{eq:process3}
\raisebox{-.8cm}{\includegraphics[scale=.48]{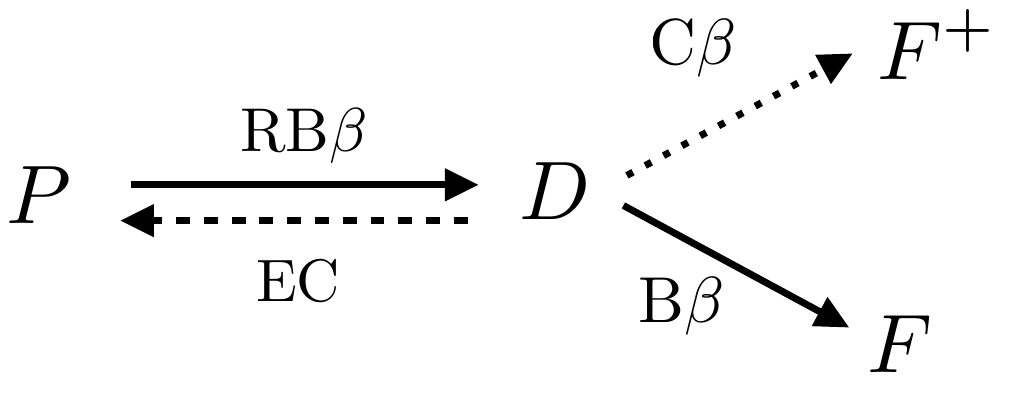}},
\end{align}
where continuous beta decay into the ionised final state, $F^+$, is also allowed. Similarly, the 3-state resonant electron capture process decay chain reads
\begin{align}\label{eq:process4}
\raisebox{-.8cm}{\includegraphics[scale=.48]{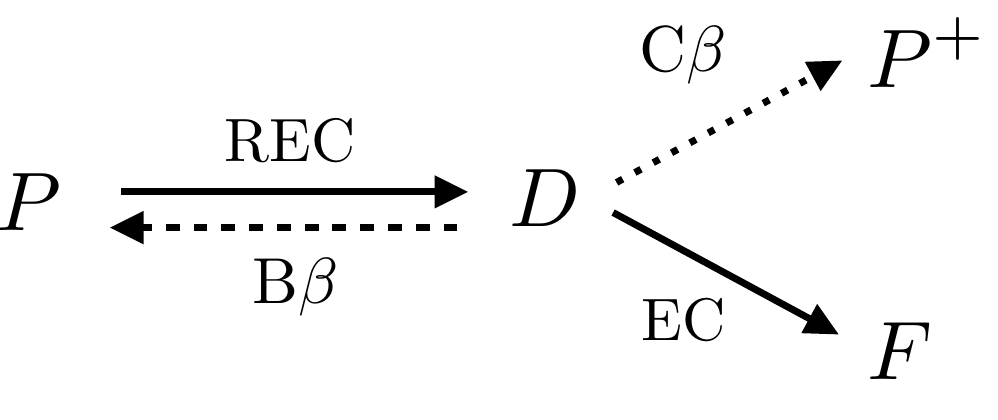}}.
\end{align}
The presence of a third state modifies the branching ratios for the decay back to the initial state,
\begin{equation}\label{eq:3state1}
    \mathcal{B}_{DP} = \begin{cases}
        \chi\mathcal{B}_\mathrm{EC}, &\quad \mathrm{RB}\beta, \\[3pt]
        \displaystyle{\frac{\chi(1-\mathcal{B}_\mathrm{EC})}{1+K(Z,Q)}}, &\quad \mathrm{REC},
    \end{cases}
\end{equation}
where $\mathcal{B}_\mathrm{EC}$ is the branching ratio for the resonance to decay by electron capture, widely available in nuclear data tables~\cite{iaeachart}. Further, the branching ratios for the daughter ion to decay into the final state are
\begin{equation}\label{eq:3state2}
    \mathcal{B}_{DF} = \begin{cases}
        \displaystyle{\frac{1-\mathcal{B}_\mathrm{EC}}{1+K(Z+2,Q_{DF})}}, &\quad \mathrm{RB}\beta, \\[3pt]
        \mathcal{B}_\mathrm{EC}, &\quad \mathrm{REC}\,, \\
    \end{cases}
\end{equation}
where $Q_{DF}$ is the $Q$-value for the $D\to F$ transition and one has to evaluate the ratio of continuous and bound beta decay widths $K(Z,Q_f)$ for $n_f=2$ in \eqref{eq:3state1} and $n_f=1$ in \eqref{eq:3state2}. See appendix~\ref{App:bbdecay} for details.  
For small quality factors $R_\tau\ll 1$, we find for the number of final state ions on the beam,
\begin{align}\label{eq:NF3statesapprox}
    N_F(x)&= N_0 R_\tau \mathcal{B}_{DF} (x + e^{-x} - 1) + \mathcal{O}(R_\tau^2)
\end{align}
whilst the number of resonance states on the beam evolves according to \eqref{eq:xySignal0}.  The full expression can be found in appendix~\ref{App:beamstates}. 

The advantages of using a 3-state system become apparent here, as \eqref{eq:NF3statesapprox} is a monotonically increasing function with $x$. The full expression~\eqref{eq:NF3states} satisfies 
\begin{equation}
    \lim_{x\to{\infty}}N_F(x)=\frac{N_0 \mathcal{B}_{DF}\chi}{\chi-\mathcal{B}_{DP}} \gg N_0 R_\tau,
\end{equation}
such that unlike 2-state systems, there is no equilibrium at $x \simeq 1$ and given an infinite amount of time, a significant fraction of the initial state ions will be converted to a stable final state. The remainder will end up in either the $F^+$ or $P^+$ states, which are ejected from the beam. 

The inclusion of the third state typically leads to a reduction in $\mathcal{B}_{DP}$ and subsequently the quality $R_\tau$ is reduced compared to a 2-state system with the same $Q$-value. In contrast to 2-state systems, there are no clear disadvantages in using processes with daughter states that have as short a lifetime as possible and an experiment with a 3-state system therefore retains a fraction of the signal even if the beam is slowed down.
Additionally, in the large $x$ limit for 3-state systems the lack of an equilibrium value allows $N_F$ to far exceed $N_D$ for an equivalent 2-state system.
This leads to a remarkable result: we can achieve a resonantly enhanced neutrino cross sections with a \emph{stable, clean} signal.

The goal is then twofold; to maximise the quality $R_\tau$ such that the number of signal states on the beam increases as quickly as possible and to find resonance states with the smallest possible lifetime $\gamma \tau_D$, such that a larger $x$ can be achieved with minimal runtime. Both of these can be achieved by minimising the $Q$-value, as $R_\tau \propto Q^{-2}$ and $\gamma \propto Q^{-1}$. We further note that the ratio of bound beta decay to continuous beta decay is larger for smaller $Q$-values, leading to an enhanced branching ratio $\mathcal{B}_{DP}$ for resonant electron capture, see appendix~\ref{App:bbdecay} for details. Since the $Q$ value also determines the beam energy required for the experiment, it is the most important parameter for both 2 and 3-state systems.

\subsection{Excited states}
The bottleneck for a realistic experiment with resonant states remains the $Q$-value, which has to be sufficiently low for an accelerator experiment to probe neutrino masses at the eV scale or below. In principle, a lower $Q$-value can be achieved for any given target by considering excited nuclear states. Exciting the initial state increases its effective mass, reducing the effective $Q$-value for neutrino capture in \eqref{eq:QREC} and \eqref{eq:QRBbeta}. The exact reduction in $Q$ that can be achieved depends strongly on the excited states that exist for a given target, but can be anywhere from $\mathcal{O}(10\,\mathrm{keV})$ to $\mathcal{O}(1\,\mathrm{MeV})$. We do not speculate about the experimental challenges related to exciting and accelerating the ions here, but show how transitions with excited states could lead to higher rates than resonant neutrino capture experiments with ground state ions. 
Exciting the initial states $P^*$ comes at the cost of making them unstable, with a lifetime $\tau_{P^*}$. We discuss resonant neutrino capture processes with excited states using both 2- and 3-state systems here. First, we consider a 2-state system analogous to \eqref{eq:process1} with an excited state $P^*$ that can undergo resonant bound beta decay by capturing a neutrino, 
\begin{align}\label{eq:process1ex}
\raisebox{-1.6cm}{\includegraphics[scale=.48]{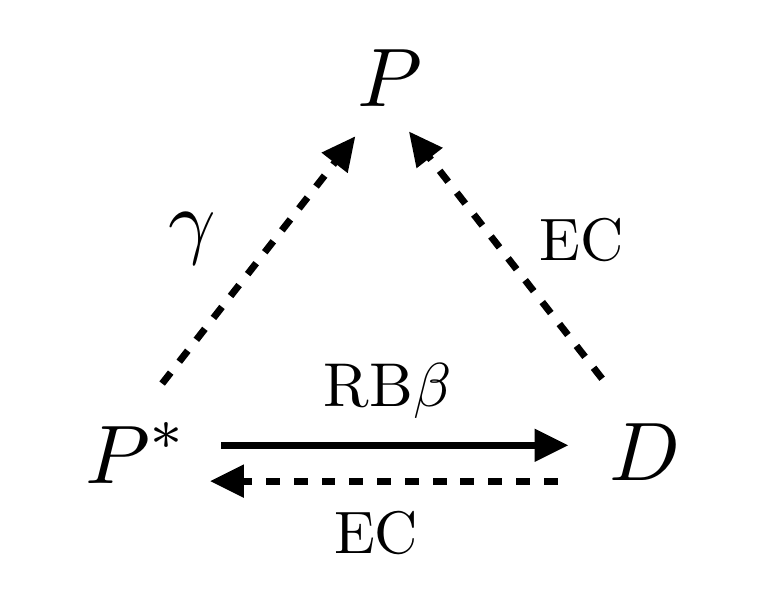}}.
\end{align}
The excited state $P^*$ is accelerated on a beam and can either produce the signal state $D$ through bound beta decay or decay back into its ground state $P$ by emission of photon. The daughter state $D$ can capture an electron and decay back into the initial state $P^*$ or its ground state $P$. 
 
Alternatively, analogous to \eqref{eq:process2}, the neutrino can be captured in an resonant electron capture process by an excited state,
\begin{align}\label{eq:process2ex}
\raisebox{-1.6cm}{\includegraphics[scale=.48]{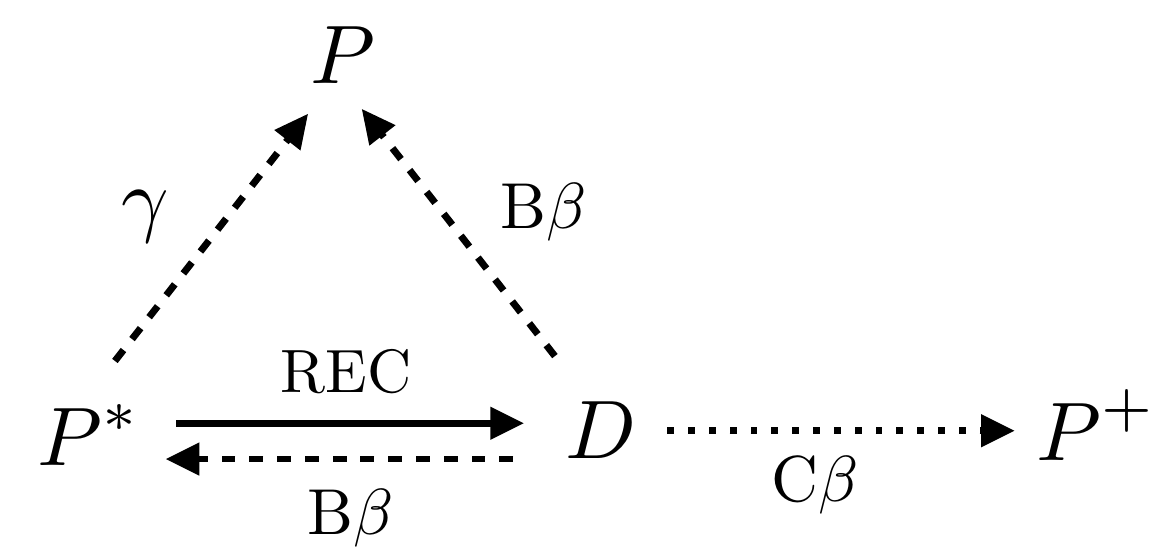}},
\end{align}
where the daughter ion can bound beta decay into both the initial state $P^*$ or its ground state $P$, or continuous beta decay to $P^+$. Since the beam energy needs to be tuned for resonant neutrino capture with $P^*$, the larger $Q$-value of $P$ renders it unable to capture relic neutrinos at the same beam energy.  

Following the procedure outlined in appendix~\ref{App:beamstates}, we can derive the number of signal state particles $D$ on the beam for the system ~\eqref{eq:process1ex}. In the approximation $R_\tau\ll 1$ we find in terms of the lifetime ratio of the daughter and excited state, $\eta=\tau_D/\tau_{P^*}$,
%
%
\begin{align}
    N_D(x) &= \frac{N_0 R_\tau}{\eta-1} (1-e^{-x(\eta-1)})e^{-x} + \mathcal{O}(R_\tau^2) \\
    &= \begin{cases}
    \frac{N_0 R_\tau}{\eta} (1- e^{-\eta x}) e^{-x}, \quad &\eta \gg 1 \\[3pt]
    N_0 R_\tau (1- e^{-x}), \quad &\eta \ll 1
    \end{cases}\,.
\end{align}
This equation reproduces \eqref{eq:xySignal0} in the long excited state lifetime limit $\eta \ll 1$, and in this case the excited state system retains the same properties previously discussed in Section~\ref{sec:2stateGS}. However, it has drastically different properties in the short parent lifetime, $\eta \gg 1$, regime.  For large $\eta$, the maximum population of signal states on the beam is obtained for a runtime 
\begin{equation}\label{eq:xmax2}
    x_{\mathrm{max}} 
    \simeq \frac{1}{\eta} \ln\eta\,, \qquad \eta \gg 1\,,
\end{equation}
which is significantly smaller than for the ground state system where $x_\mathrm{max}\simeq 1$. A very short lifetime of $P^*$ rapidly depletes the beam of excited initial states $P^*$, leaving a beam which consists entirely of ground states $P$ and daughter states $D$. For $x> x_\mathrm{max}$, the number of resonance states start to decay according to 
\begin{equation}
    N_D(x > x_\mathrm{max}) \simeq \frac{N_0 R_\tau}{\eta} e^{-x},
\end{equation}
such that for $x_\mathrm{max} < x < 1$, the number of signal states $D$ remains close to its maximal value, and the signal states only begin to rapidly decay for a runtime $x \gtrsim 1$.

The maximal number of signal states in an experiment with an excited initial state can then be compared to the maximal signal state population in an experiment designed to capture neutrinos with its ground state isomer as described in Section~\ref{sec:2stateGS}. The excited state system is superior to the corresponding ground state system if $x_\mathrm{max} < x < 1$ and their respective quality factors fulfil
\begin{equation}
    \frac{1}{\eta} R_{\tau,\mathrm{ES}} > R_{\tau,\mathrm{GS}}\,,
\end{equation}
where GS and ES refer to the ground state and excited state system quantities respectively.
Neglecting different branching ratios and spin factors, this translates into a condition on the $Q$-values, 
\begin{equation}
\sqrt{\eta} \, Q_\text{ES}\lesssim Q_\text{GS}\,,
\end{equation}
which implies that excited state systems only present a better choice for very small $Q$-values or $\eta$. \emph{If} it is possible to re-excite $P$ states on the beam, multiple runs with the excited initial states for a runtime $x_\mathrm{max}$ each would have the potential to produce a sizable population of $D$ states in the same time needed to reach the maximum signal state population for a ground state system at $x\simeq 1$. In this case the experiment with the excited state system is superior to the equivalent experiment performed with a ground state system if
\begin{equation}\label{eq:Qexcited}
       \frac{1}{\ln\eta} R_{\tau,\mathrm{ES}} >  R_{\tau,\mathrm{GS}}\,\implies \sqrt{\ln\eta}\, Q_\text{ES}\lesssim Q_\text{GS}\,,
\end{equation}
where we set $x=1$ and neglected different branching ratios and spin factors. As in the case of 2-state systems with a ground state parent ion, the signal state is unstable and will decay.

As before, one can produce stable signal states in a 3-state system with an excited initial state. For resonant bound beta decay, the excited 3-state system can be written in analogy to \eqref{eq:process3} as 
\begin{equation}\label{eq:process3ex}
\raisebox{-.4cm}{\includegraphics[scale=.48]{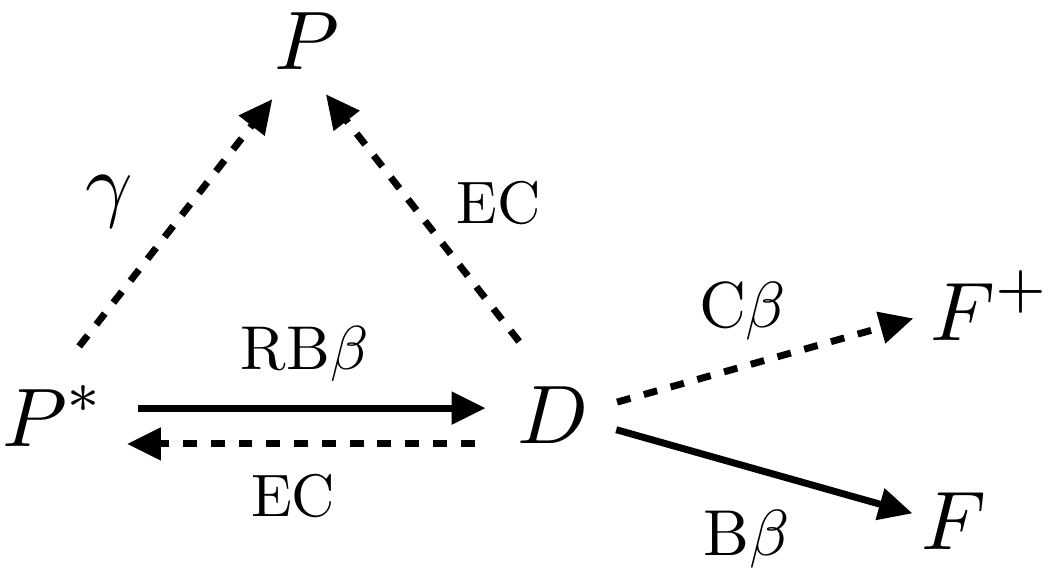}}\,,
\end{equation}
and in the case of resonant electron capture in analogy to \eqref{eq:process4} as 
\begin{equation}\label{eq:process4ex}
\raisebox{-.6cm}{\includegraphics[scale=.48]{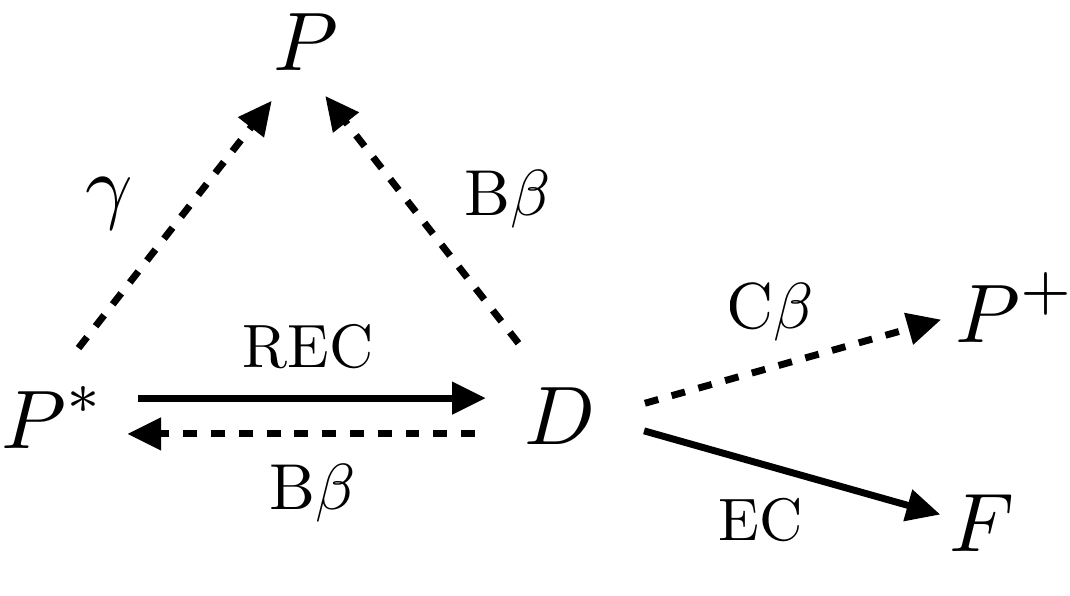}}\,.
\end{equation}
The number of stable signal states $F$ on the beam at time $x$ is expected to follow
\begin{align}
    N_F(x) 
    &= \frac{N_0 \mathcal{B}_{DF} R_\tau}{\eta} \left(1 + \frac{\eta e^{-x} - e^{-\eta x}}{1 - \eta}\right) + \mathcal{O}(R_\tau^2) \\
    &\simeq
    \begin{cases}
    \frac{N_0 \mathcal{B}_{DF} R_\tau}{\eta} (1 - e^{-x}),&\quad \eta \gg 1, \\[3pt]
    N_0 \mathcal{B}_{DF} R_\tau (x+e^{-x}-1),&\quad \eta \ll 1,
    \end{cases}
\end{align}
where the small $\eta$ limit reproduces~\eqref{eq:NF3statesapprox} as expected.

%
%
In contrast to the ground state system however, the limiting value for the number of signal states on the beam is given by
\begin{equation}
    \lim_{x\to{\infty}}N_F(x) = \frac{N_0  \mathcal{B}_{DF} R_\tau}{\eta+ R_\tau (1-\mathcal{B}_{DP})}.
\end{equation}
For short excited state lifetimes with large values of $\eta$, the number of signal states saturates at $x\simeq 1$ with $N_F(x>1)\ll N_0$. Therefore by using excited $P$ states, we have lost the ability for 3-state systems to convert a significant fraction of the beam at large $x$, although we retain a clean, stable signal. 

For a repeated excitation of the initial state $P \to P^*$ the condition for which the excited 3-state system produces more signal states than an equivalent ground state system in a given time $x$ is given by \eqref{eq:Qexcited}.\\

\begin{figure}
    \centering
    \includegraphics[width=\linewidth]{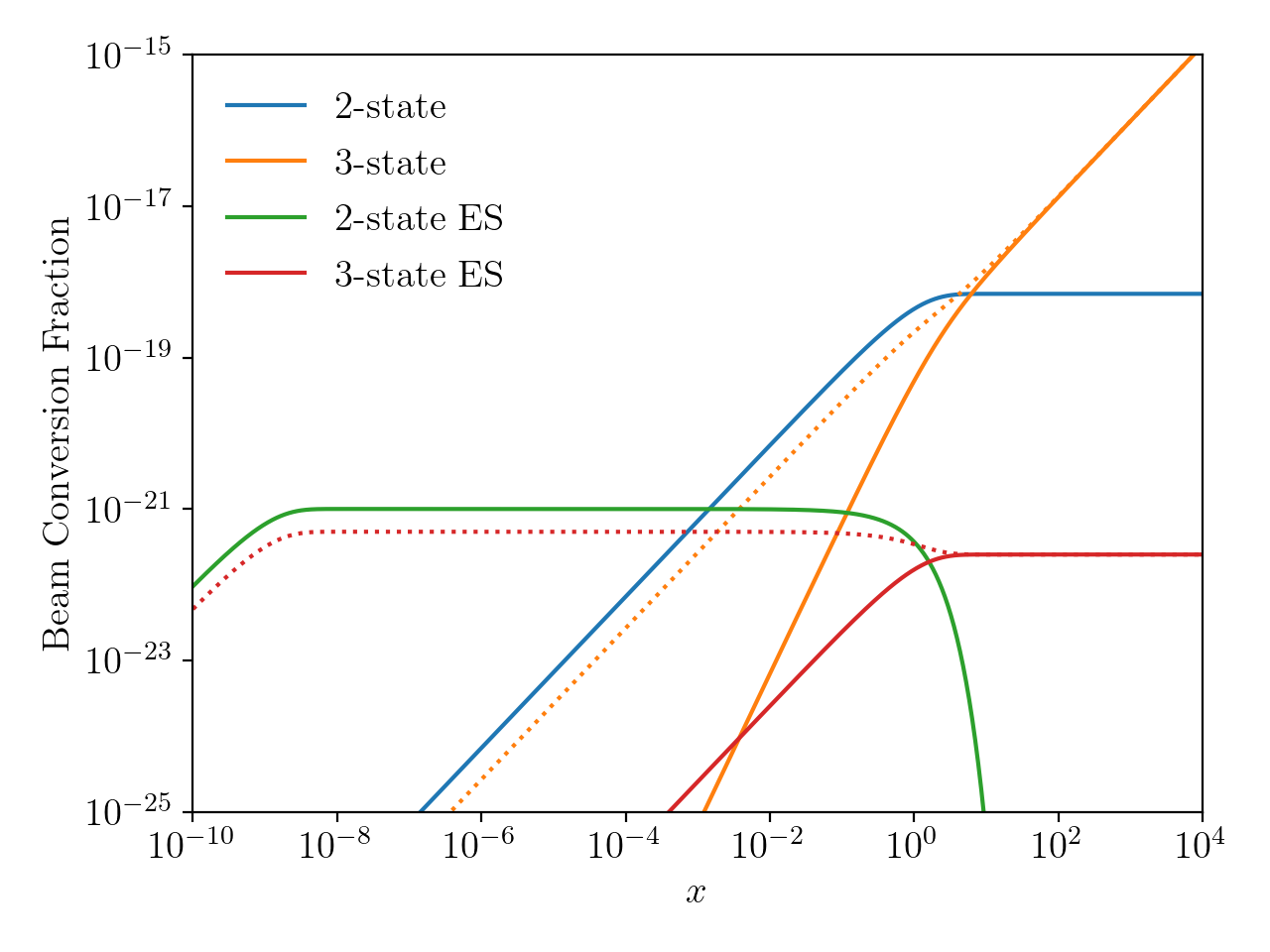}
    \caption{Fraction of parent ions converted to signal ($D$ or $F$ states) for different resonant systems, all with $\chi = 1$, $J_D = J_P$, $m_\nu = 0.1\,\mathrm{eV}$, $Z = 50$, $A = 100$, $\delta_b=0$ and $\mathcal{B}_{EC} = 0.5$ where appropriate. For the ground state (GS) systems, $Q=100\,$keV, whilst the excited state (ES) systems use $Q = 0.1\,$keV and $\eta = 10^{9}$. Finally, for the 3-state systems we plot $N_F$ using a solid line, and $N_F + N_D$ using a dotted line.}
    \label{fig:comparisonAll}
\end{figure}

In Figure~\ref{fig:comparisonAll} we compare the expected fraction of a beam of ions that is converted into the signal state as a function of $x$. The different contours correspond to the resonant electron capture process in the 2-state system \eqref{eq:process2}, shown in blue, the 3-state system \eqref{eq:process4}, shown in orange, and the analogous transition with excited initial states shown in green in the case of the 2-state system in \eqref{eq:process2ex} 
and in red for the 3-state system in \eqref{eq:process4ex}. 
The resonant bound beta decay process leads to slightly different conversion rates, due to different contributions from continuous beta decay to the total decay width. 
Since we generally expect smaller $Q$-values for excited states we have fixed $Q=100$ keV for the ground state systems in Figure~\ref{fig:comparisonAll} and  $Q=0.1$ keV for the excited states. 

Focusing first on the ground state systems, Figure~\ref{fig:comparisonAll} shows that the 2-state system reaches its maximal population of the signal state at $x\simeq 1$. Beyond that point the beam has reached equilibrium between the initial and the resonant state. Instead, for the 3-state system the final state is stable and in principle gets populated until the whole beam is either converted or ejected. As a result, 3-state systems can achieve substantially larger beam conversion rates, although this process becomes less efficient for large $x$.
The point at which the 3-state system beam conversion fraction exceeds the 2-state system is visible at $x\simeq 1$.


The main advantage of the systems with excited states is that it can achieve equilibrium at $x\sim (1/\eta) \ll 1$,  much faster than ground state systems. Multiple runs of excited 2-state systems can then have cumulative beam conversion factors that are orders of magnitude larger compared to ground state systems. The experimental realisation of these systems represents an extraordinary challenge. It requires to keep repopulating a beam with short-lived excited states and read out the signal before it is lost at $x\simeq 1-10$. A 3-state system with an excited state overcomes this problem by populating a stable final state. In contrast to the 3-state system with an initial ground state, the experiment can likely operate at substantially smaller values of $Q$. As shown in Figure~\ref{fig:comparisonAll} the maximum number of conversions is smaller compared to ground state systems, because the excited states have the additional route to decay back into the ground state. 



The results shown in Figure~\ref{fig:comparisonAll} have to be compared with the expected rate at PTOLEMY~\eqref{eq:PtolemyRate}, which corresponds to a conversion factor of $2\times 10^{-25}$ of the target tritium atoms per year. For comparison, for a runtime of $t=10 \gamma \tau_D$, the conversion rate for the example 2-state system in Figure~\ref{fig:comparisonAll} would be $\sim 10^{-19}$.

 \section{Real world examples}\label{sec:realworld}
 
 \begin{table*}[]
\renewcommand{\arraystretch}{1.5}
    \centering
    \begin{tabular}{|c|c|c|c|c|c|c|c|}
    \hline
    System & $Q\,$(keV) & $E/A\,\mathrm{(TeV)}$ & \begin{tabular}{c} $\mathcal{B}_{DP}$ \\ $\mathcal{B}_{DF}$ \\ \end{tabular} & $x\, (t=1\,\mathrm{y})$ & \begin{tabular}{c}$N_D(x)/N_0$ \\ $N_F(x)/N_0$ \end{tabular} & $\log_{10}\eta$ & $\frac{x}{x_\mathrm{max}}\frac{N_D(x_\mathrm{max})}{N_0}$ \\
    \hline\hline
    ${^{120}\mathrm{Sn}^{*}}(2482\,\mathrm{keV})\xrightarrow[]{\mathrm{RB}\beta}{^{120}\mathrm{Sb}}$ & 169.66 & $1.58\times 10^{3}$ & 1 & 0.014 & $4.29\times10^{-28}$ & 7.91 & $2.72\times10^{-23}$ \\
    \hline
    ${^{100}\mathrm{Mo}}\xrightarrow[]{\mathrm{RB}\beta}{^{100}\mathrm{Tc}} \xrightarrow[]{\mathrm{B}\beta} {^{100}\mathrm{Ru}}$ & 152.68 & $1.42\times10^3$ & \begin{tabular}{c} $3\times10^{-5}$ \\ 0.254 \end{tabular} & 0.916 & \begin{tabular}{c} $1.18\times10^{-23}$ \\ $1.58\times10^{-24}$ \end{tabular} & - & - \\
    \hline
    ${^{106}\mathrm{Cd}}\xrightarrow[]{\mathrm{REC}}{^{106}\mathrm{Ag}} \xrightarrow[]{\mathrm{EC}} {^{106}\mathrm{Pd}}$ & 214.08 & $1.99\times10^3$ & \begin{tabular}{c} 0.003 \\ 0.995 \end{tabular} & $7.11\times10^{-3}$ & \begin{tabular}{c} $7.08\times10^{-24}$ \\ $2.51\times10^{-26}$ \end{tabular} & - & - \\
    \hline
    ${^{165}\mathrm{Ho}}\xrightarrow[]{\mathrm{RB}\beta}{^{165}\mathrm{Er}}$ & 322.16 & $3.00\times10^3$ & 1 & $1.82\times10^{-4}$ & $6.69\times10^{-24}$ & - & - \\
    \hline
    ${^{157}\mathrm{Gd}}\xrightarrow[]{\mathrm{RB}\beta}{^{157}\mathrm{Tb}}$ & 10.95 & 101.95 & 0.999 & $8.91\times10^{-8}$ & $3.78\times10^{-24}$ & - & - \\
    \hline
    ${^{179}\mathrm{Hf}}\xrightarrow[]{\mathrm{RB}\beta}{^{179}\mathrm{Ta}}$ & 41.44 & 385.89 & 1 & $9.22\times10^{-7}$ & $2.19\times10^{-24}$ & - & - \\
    \hline
    ${^{71}\mathrm{Ga}}\xrightarrow[]{\mathrm{RB}\beta}{^{71}\mathrm{Ge}}$ & 222.50 & $2.07\times10^3$ & 1 & $9.98\times10^{-6}$ & $5.13\times10^{-25}$ & - & - \\
    \hline
    ${^{121}\mathrm{Sb}}\xrightarrow[]{\mathrm{REC}}{^{121}\mathrm{Sn}}$ & 430.94 & $4.02\times10^3$ & 0.351 & $5.22\times10^{-5}$ & $3.35\times10^{-25}$ & - & - \\
    \hline
    ${^{64}\mathrm{Zn}}\xrightarrow[]{\mathrm{REC}}{^{64}\mathrm{Cu}} \xrightarrow[]{\mathrm{EC}} {^{64}\mathrm{Ni}}$ & 588.12 & $5.48\times10^3$ & \begin{tabular}{c} 0.047 \\ 0.615 \end{tabular} & $8.13\times10^{-5}$ & \begin{tabular}{c} $1.69\times10^{-25}$ \\ $4.22\times10^{-30}$ \end{tabular} & - & - \\
    \hline
    ${^{104}\mathrm{Ru}^{*}}(988\,\mathrm{keV})\xrightarrow[]{\mathrm{RB}\beta}{^{104}\mathrm{Rh}} \xrightarrow[]{\mathrm{B}\beta} {^{104}\mathrm{Pd}}$ & 126.73 & $1.18\times10^3$ & \begin{tabular}{c} $10^{-5}$ \\ $0.007$ \end{tabular} & 0.408 & \begin{tabular}{c} $1.18\times10^{-36}$\\ $4.15\times10^{-39}$ \end{tabular} & 12.73 & $1.32\times10^{-25}$ \\
    \hline
    ${^{3}\mathrm{He}}\xrightarrow[]{\mathrm{REC}}{^{3}\mathrm{H}}$ & 18.58 & 174.00 & 0.012 & $3.03\times10^{-7}$ & $5.36\times10^{-26}$ & - & - \\
    \hline
    ${^{171}\mathrm{Yb}}\xrightarrow[]{\mathrm{REC}}{^{171}\mathrm{Tm}}$ & 154.66 & $1.44\times10^3$ & 0.940 & $2.34\times10^{-7}$ & $4.68\times10^{-26}$ & - & - \\
    \hline
    ${^{63}\mathrm{Cu}}\xrightarrow[]{\mathrm{REC}}{^{63}\mathrm{Ni}}$ & 74.90 & 696.91 & 0.626 & $9.17\times10^{-9}$ & $2.60\times10^{-27}$ & - & - \\
    \hline
    ${^{107}\mathrm{Ag}}\xrightarrow[]{\mathrm{REC}}{^{107}\mathrm{Pd}}$ & 57.72 & 537.22 & 0.970 & $1.85\times10^{-13}$ & $8.22\times10^{-31}$ & - & - \\
    \hline
    \end{tabular}
    \caption{Example targets for resonantly capturing cosmic neutrinos, listed alongside their capture threshold $Q$, the beam energy per nucleon required to hit the resonance $E/A$, branching ratios to decay to the initial/final state $\mathcal{B}_{DP/DF}$, and the beam conversion fractions after one year of runtime, $N_{D/F}/N_0$. For the definitions of the quantities $x$, $\eta$ and $x_{\mathrm{max}}$, please refer to the text. Note that the RB$\beta$ initial states have 0 electrons, whilst the 2 and 3-state REC systems have 1 and 2 initial electrons respectively. When calculating $N_D$ and $N_F$ we use $f_{c,i} = 1$, $\delta_b = 0$ and $\delta_\nu = 0.00334$, corresponding to $m_\nu = 0.1\,\mathrm{eV}$ and $T_\nu = T_\nu^0$. Input values are taken from~\cite{iaeachart} and~\cite{nist}.}
    \label{tab:rwSystems}
\end{table*}

 Whilst promising in theory, the efficacy of a resonant neutrino capture experiment will depend strongly on whether or not an ion system with the right properties exists. In the previous section we have shown that the properties necessary to maximise the conversion rate independent of the process are a small $Q$-value, a short resonance lifetime and an $\mathcal{O}(1)$ branching ratio for the resonance to decay to both the initial and final states. Here we compile a list of several ion systems with these desired properties in Table~\ref{tab:rwSystems}. We emphasise that these examples demonstrate that such isotopes exist in nature, but our list is not extensive. It is likely that better systems using different ions or more esoteric systems like ionised molecules can be identified. 
 
 
 In Table \ref{tab:rwSystems} we list a number of ion systems and the number of neutrinos captured per target ion after a 1 year experimental runtime.
 The first two rows are examples of excited initial states, for which we use the last two columns to denote the ratio of lifetimes between the initial and resonant states as well as the event rate that could be achieved if the initial state is re-excited $x/x_\text{max}$ times, respectively.   
 
 We find that in general the beam conversion fractions are limited by $x \ll 1$, resulting from the relatively long weak decay lifetimes that are extended by large Lorentz factors $\gamma = Q/m_\nu$. Additionally, many of the systems listed suffer from prohibitively large beam energy requirements, of order $(E/A) > 1\,$PeV per nucleon. 
 
 Despite this, there are several targets that show some promise. In particular, an experiment using the two state system ${^{157}\mathrm{Gd}}\xrightarrow[]{\mathrm{RB}\beta}{^{157}\mathrm{Tb}}$ could be performed with a beam energy of $101.95\,\mathrm{TeV}$ for $m_\nu = 0.1\,\mathrm{eV}$, which is achievable with existing proposals for a Future Circular Collider (FCC)~\cite{Benedikt:2018csr}. In addition to the states listed in Table~\ref{tab:rwSystems}, there exist states which are kinematically allowed to decay when suitably ionised but which have no reported lifetime, making their associated beam conversion fractions difficult to estimate. For example, the system ${^{107}\mathrm{Pd}}\xrightarrow[]{\mathrm{REC}}{^{107}\mathrm{Ag}}$ is a candidate with $Q=2.87\,\mathrm{keV}$ when fully ionised, whilst ${^{194}\mathrm{Os}}\xrightarrow[]{\mathrm{REC}}{^{194}\mathrm{Ir}}$ is a system with threshold $Q =4.62\,\mathrm{keV}$. These thresholds correspond to beam energies per nucleon of $26.71\,\mathrm{TeV}$ and $43.03\,\mathrm{TeV}$ respectively, within reach of an experiment not significantly larger than the LHC.

A more significant challenge posed by this experiment is getting sufficiently many ions on the beam to observe an event with a runtime of 1 year. In the aforementioned ${^{157}\mathrm{Gd}}\xrightarrow[]{\mathrm{RB}\beta}{^{157}\mathrm{Tb}}$ system, an event rate of 1 per year would require $2.65\times10^{23}$ ions on the beam, corresponding to a total beam mass of \SI{69.1}{\gram}. This number far exceeds the number of ions on the LHC beam~\cite{Bruce:2021hii}, however, as we no longer wish to collide the beam, it is difficult to compare our experiment to a traditional collider experiment. We speculate on what may be possible in simplified scenarios in Section~\ref{sec:beamLimitations}.

\section{Beam Limitations} \label{sec:beamLimitations}
Performing a resonant neutrino capture experiment to detect the C$\nu$B is fraught with difficulties. In particular, such an experiment would require accelerating a macroscopic number of particles to extremely high energies, whilst being able to precisely hit the resonance. In this section we attempt to estimate what can realistically be achieved in some simplified scenarios. 

\subsection{Neutrino mass uncertainty}
\begin{figure}
    \centering
    \includegraphics[width=\linewidth]{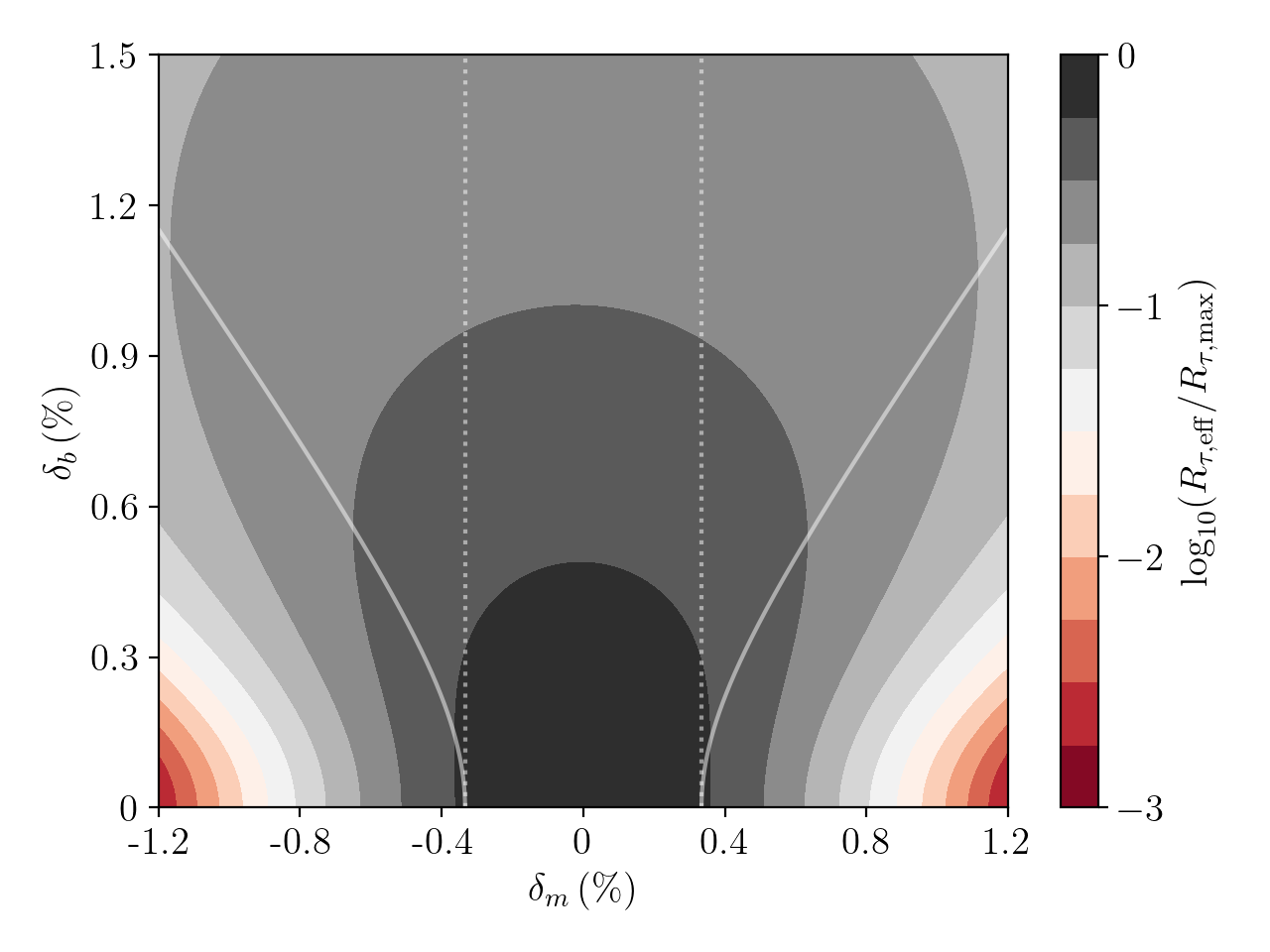}
    \caption{Effect of the neutrino mass uncertainty on $R_\tau$, with respect to its maximum value. For simplicity, we set $R_\tau = 1/\sqrt{\delta_\nu^2 + \delta_b^2}$. As before, we use $\delta_\nu = 0.00334$, corresponding to $m_{\nu,\mathrm{true}} = 0.1\,$eV and $T_\nu = T_\nu^0$. The solid white contours denote the value of $\delta_b$ that maximises $R_\tau$ at a given value of $\delta_m$, whilst the white dotted contours are located at $|\delta_m| = \delta_\nu$.}
    \label{fig:massUncertainty}
\end{figure}

As C$\nu$B neutrinos are non-relativistic, we require prior knowledge of the neutrino mass to precisely hit the resonance. However, any direct measurement of the neutrino mass will have some uncertainty which, if left unaccounted for, will lead to the ion beam being tuned to the incorrect energy. Supposing that the experiment is run with the assumption of a neutrino mass $m_{\nu,\mathrm{pred}}$ that differs from the true neutrino mass, $m_{\nu,\mathrm{true}}$, the event rate is modified to
\begin{equation} \label{eq:rateModified}
\displaystyle    \frac{R_{\mathrm{eff}}}{N_T} = \frac{R}{N_T} (1-\delta_m) e^{-\frac{\delta_m^2}{2 (\delta_\nu^2 + \delta_b^2)}},
\end{equation}
where we continue to assume a Gaussian distribution for $\widetilde{E}_\nu$ and define the signed fractional uncertainty on the neutrino mass $\delta_m = \frac{m_{\nu,\mathrm{true}} -m_{\nu,\mathrm{pred}}}{m_{\nu,\mathrm{true}}}$, whilst $R/N_T$ is given by \eqref{eq:intRate}. This results in an effective quality factor given by,
\begin{equation} \label{eq:rtauModified}
    R_{\tau,\mathrm{eff}} = R_\tau (1-\delta_m)^2 \displaystyle e^{-\frac{\delta_m^2}{2 (\delta_\nu^2 + \delta_b^2)}},
\end{equation}
where $R_\tau$ takes the form in \eqref{eq:Rtau} with $m_\nu = m_{\nu,\mathrm{true}}$. Then in order to avoid exponential suppression of the event rate, one needs to ensure that the exponent appearing in both \eqref{eq:rateModified} and \eqref{eq:rtauModified} is less than unity. Given that $\delta_\nu$ is fixed by nature and $\delta_m$ depends on the progress of neutrino mass experiments, this can only be achieved by modifying the beam momentum spread $\delta_b$. In particular, \eqref{eq:rtauModified} has a maximum when
\begin{equation}
    \delta_b = \begin{cases}
        0, \quad &|\delta_m| < \delta_\nu \\[3pt]
        \sqrt{\delta_m^2 - \delta_\nu^2}, \quad &|\delta_m| > \delta_\nu 
    \end{cases}
\end{equation}
where we have included the dependence of $R_\tau$ on $\delta_\nu$ and $\delta_b$. We show the effect of the neutrino mass uncertainty on the effective quality factor $R_{\tau,\mathrm{eff}}$ in Figure \ref{fig:massUncertainty}. We also show the contour $|\delta_m| = \delta_\nu$ using dotted lines, and we show the optimal value of $\delta_b$ for $|\delta_m| > \delta_\nu$ and a given $\delta_m$ using solid white contours. For reference, RHIC runs with a beam momentum spread $\delta_b\sim 10^{-4}$~\cite{Luo:2016hgj}.

As a consequence of $\delta_m$ being a signed quantity there is a slight asymmetry of \eqref{eq:rtauModified} in Figure \ref{fig:massUncertainty}.
This results in $R_{\tau,\mathrm{pred}}$ having a maximum at $\delta_m < 0$, suggesting that it is better to underestimate the neutrino mass in this experiment, as the higher beam energy will allow for more neutrinos in the lower energy tail of the distribution to cross the interaction threshold.  

\subsection{Number of targets}
A large number of highly charged ions on an energetic beam can damage the magnets and the beampipe through the loss of beam particles and synchrotron radiation.
Damage from beam loss is reduced by the collimator system, which prevents quenching of the magnets by removing ions from the beam halo. The performance of the collimators can be measured by the local cleaning inefficiency $\eta_C$, which is the fraction of the beam ions leaked to sensitive components of the experiment over the number of ions absorbed by the collimators per unit length~\cite{Redaelli:2016ohw}. The cleaning inefficiency needs to be low enough such that a beam with $N_T$ ions and energy $E$ does not exceed the quench limit, 
\begin{equation}
    R_q(E) = 7.8\times 10^{8}\,\mathrm{m}^{-1}\,\mathrm{s}^{-1} \left[\frac{7\,\mathrm{TeV}}{E}\right]^{\frac{3}{2}} ,
\end{equation}
where the reference value has been determined for a proton beam in~\cite{Jeanneret:1996qx} and the scaling with beam energy in~\cite{Bracco:2009zz}. 
We are interested in the maximum number of ions that can be put on a beam for a time $\tau_b$ before the magnets quench, 
\begin{align}\label{eq:quench}
    N_T&=\frac{\tau_b}{\eta_C} R_q(E)\\
    &\simeq 2.74\times 10^{15} \left[\frac{\tau_b}{100\,\mathrm{h}}\right] \left[\frac{10^{-5}\,\mathrm{m}^{-1}}{\eta_C}\right] \left[\frac{10\,\mathrm{TeV}}{E}\right]^{\frac{3}{2}}.\notag
\end{align}
To derive this number we have used the cleaning inefficiency achieved by the LHC collimator system~\cite{Redaelli:2016ohw} and assume a beam lifetime of $\tau_b=100\,$h~\cite{Benedikt:2018csr}.
An experiment designed to observe relic neutrinos is expected to have fewer beam losses because no beam collisions take place.
In order to significantly exceed the limit \eqref{eq:quench} however, substantial progress on increasing both beam stability and cleaning efficiency is required. Additionally, the scaling with the beam energy reinforces the need for small $Q$-values.  
Even if ion losses from the beam can be controlled the large number of ions necessary to observe resonant neutrino capture events emit a significant amount of synchrotron radiation, which may also cause damage to the experiment. 
This is due to the heating of the beam pipe through absorption and the emission of molecules and electrons from the beam pipe walls. Here we put a very conservative upper limit on the number of target ions that can be accelerated before the heat induced by the synchrotron radiation melts the beam pipe, the so-called ``death beam limit".

We consider a simplified scenario in which the current in the beampipe is constant, allowing us to neglect ohmic heating. Such an experiment would have a beam of constant number density, as opposed to bunches of particles. The heating of the beampipe walls will then be exclusively due to incident synchrotron radiation, which is emitted with total power
\begin{equation}
    P_\mathrm{tot}= \frac{2\alpha I^2 \gamma^4}{3 R_{\mathrm{c}}^2}N_T
\end{equation}
for an ion accelerator ring with radius $R_\mathrm{c}$ and target ions with degree of ionisation $I$. The synchrotron radiation is focused in a cone with opening angle $\theta \simeq 2/\gamma$, such that it strikes the wall of the beampipe with power per unit area
\begin{equation}
    P_{\mathrm{area}} \simeq \frac{\gamma}{4\sqrt{2}\pi R_\mathrm{c}^{\frac{3}{2}} \sqrt{r}} P_\mathrm{tot} = \frac{\alpha I^2 \gamma^5}{6\sqrt{2}\pi R_\mathrm{c}^{\frac{7}{2}}\sqrt{r}}N_T,
\end{equation}
where $r$ denotes the beampipe radius.

We next assume that after some time the inner wall of the beampipe reaches an equilibrium temperature $T_\infty$, which should be chosen to avoid damage to the experiment, whilst the outer walls of the beampipe are in contact with some coolant held at temperature $T_c$. Using Fourier's law and considering only radial heat flow, the heat conducted away from the beampipe per unit area will then be given by
\begin{equation}
    q_\mathrm{con} = \kappa_\mathrm{con}\frac{dT}{dl} = \frac{\kappa_\mathrm{con}}{\Delta}(T_\infty - T_c),
\end{equation}
for a beampipe of thickness $\Delta$ with constant thermal conductivity $\kappa_{\mathrm{con}}$. The loss of heat per unit area from the beampipe is given by
\begin{equation}
    q_\mathrm{rad} = \varepsilon \sigma (T_\infty^4 + T_c^4)\,,
\end{equation}
 where $\varepsilon$ and $\sigma$ are the emissivity of the pipe walls and the Stefan-Boltzmann constant respectively. The last step is to solve the equilibrium condition 
\begin{equation}
    a P_\mathrm{area} = q_\mathrm{con} + q_\mathrm{rad} \equiv q_\mathrm{out}(T_\infty, T_c)
\end{equation}
for $N_T$, where $a$ $\in [0,1]$ is the absorptance that accounts for incomplete absorption of synchrotron radiation by the beampipe. We find that the maximum number of ions that can be accelerated for equilibrium temperature $T_\infty$ is then given by
\begin{equation}\label{eq:maxtargets}
    N_T = \frac{6\sqrt{2}\pi R_\mathrm{c}^\frac{7}{2}\sqrt{r}}{\alpha  I^2 a} \left(\frac{m_\nu}{Q}\right)^5 q_\mathrm{out}(T_\infty, T_c),
\end{equation}
where we have used $\gamma = Q/m_\nu$. This further emphasises that a small $Q$-value is the most important parameter for the choice of target ions in this experiment. 

Using \eqref{eq:maxtargets} we can derive a crude estimate for an example of a $27\,\mathrm{km}$ ion accelerator ring, with beampipe radius $r = 5\,\mathrm{cm}$ and thickness $\Delta = 1\,\mathrm{cm}$, made out of aluminium for which $\kappa_\mathrm{con} = 240\,$\SI{}{\watt\per\metre}$\,\mathrm{K}^{-1}$, $\varepsilon = 0.03$ and $a = 0.1$. We further set the coolant temperature to zero, $T_c = 0\,\mathrm{K}$ and the equilibrium temperature just below the melting point of aluminium at $T_\infty = 900\,\mathrm{K}$. This gives an upper limit on the number of ions that can be put on the beam,
\begin{equation}
    N_T \simeq 3.88\times 10^{17} \left[\frac{50}{Z}\right]^2 \left[\frac{m_\nu}{0.1\,\mathrm{eV}}\right]^5  \left[\frac{1\,\mathrm{keV}}{Q}\right]^5,
\end{equation}
assuming a fully ionised beam such that $I = Z$. This example would correspond to a beam energy per nucleon $E/A\simeq 10$ TeV with $m_\nu = 0.1\,\mathrm{eV}$ and $Q = 1\,\mathrm{keV}$. In order to achieve the much larger numbers necessary for the real world examples presented in Section~\ref{sec:realworld}, significant additional measures such as beam pipe cooling and photon stoppers are needed.

\subsection{Non-constant charge radius}

Different members of the isobars in the 2 and 3-state systems considered for resonant neutrino capture can have different charge radii. If the difference in the charge radii is larger than the beampipe radius, the resonance states are eventually lost from the beam due to collisions with the beampipe walls. For 3-state systems, we have the further constraint that the parent and final state charge radii must not differ by more than the beampipe radius. We can express these conditions as
\begin{align}
    |R_{\mathrm{c},P} - R_{\mathrm{c},D}| < r, \quad &\text{2 and 3-state systems} \label{eq:23condition}\\
    |R_{\mathrm{c},P} - R_{\mathrm{c},F}| < r, \quad &\text{3-state systems} \label{eq:3condition}
\end{align}
where $P$, $D$ and $F$ denote the parent, daughter and final state ions, respectively. For magnetic field strength $B$ the charge radius for state with 3-momentum magnitude $p_i$ is given by
\begin{equation}
    R_{\mathrm{c}} = \frac{p_i}{2 B I\sqrt{\pi \alpha}}.
\end{equation}
If we consider the C$\nu$B neutrinos to be at rest in the lab frame, then by conserving momentum we trivially recover that the resonance momentum $p_r = p$ such that the charge radii for the parent and daughter states are equal and condition \eqref{eq:23condition} is always satisfied.

The scenario for 3-state systems is somewhat more complicated, as the decaying resonance will impart some fraction of its momentum to the outgoing neutrino. Further, the final state momenta in the $1\rightarrow2$ decay system are not fully determined, and we can only put bounds on the magnitude of the final state momentum $p_f$. From energy and momentum conservation we have
\begin{align}
    &E_r = E_f + E_{\nu,f} \label{eq:energyConservation} \\
    &p_{\nu,f}^2 = p^2 + p_f^2 -2p\,p_f\cos(\varphi), \label{eq:momentumConvservation}
\end{align}
where $\varphi$ is the angle between the initial and final state ions. We can then use \eqref{eq:momentumConvservation} to derive the constraint
\begin{equation}
    \left| \frac{p^2 + p_f^2 - p_{\nu,f}^2}{2p\,p_f} \right| \leq 1,
\end{equation}
where $p_{\nu,f}$ is obtained from \eqref{eq:energyConservation}. Expanding to leading order in small quantities we find
\begin{equation}
    \frac{|p_f - p|}{p} = \frac{|R_{\mathrm{c},P} - R_{\mathrm{c},F}|}{R_{\mathrm{c},P}}\lesssim \frac{2 Q_{DF}}{M},
\end{equation}
where $Q_{DF}$ is the $Q$-value of the $D\rightarrow F$ decay. It is important to note that typically $R_{\mathrm{c},F} < R_{\mathrm{c},P}$. For 3-state systems, we are therefore limited to using systems with $Q_{DF}$ satisfying
\begin{equation}
    Q_{DF} < \frac{M}{2} \frac{r}{R_{\mathrm{c},P}}.
\end{equation}
The different charge radii could also serve as a method for extracting $F$ states without stopping the beam. 

\section{Conclusions}\label{sec:conclusions}
 Attempts at detecting the C$\nu$B are fraught with a series of seemingly insurmountable challenges, namely the tiny neutrino interaction cross sections and huge energy thresholds required for interactions to take place. 
 Neutrino capture on beta decaying nuclei has no energy threshold, and represents the most promising method of detecting relic neutrinos today. However, the cross section for neutrino capture is suppressed by $G_F^2$, and the natural beta decay produces an unavoidable background.
 
 Accelerating targets on a beam can enhance the neutrino capture cross section whilst also allowing for neutrino capture on radioactively stable targets.
 We have performed a calculation of the expected neutrino capture rates for a number of experimental realisations of this idea and discussed the challenges associated with accelerating a large number of ions.
 
 We first explored the possibility of accelerated tritium ions, the  \textit{PTOLEMY-on-a-beam} experiment, aiming to exploit the quadratic growth of the capture cross section with energy. While the neutrino capture cross section grows substantially above \SI{}{\peta\electronvolt} beam energies,
 this experiment would have a huge beta decay background. Utilising the inverse process $^3\mathrm{He}\rightarrow{^{3}\mathrm{H}}$ on a beam overcomes this problem, but still requires beam energies of a PeV or more to substantially increase the neutrino capture cross section.
 
 Our main finding is that resonant neutrino capture with an accelerator experiment allows for significantly larger neutrino capture rates, and provides a distinct final state. For a neutrino mass of $m_\nu\gtrsim 0.1\,\mathrm{eV}$, we have shown that an appropriate choice of targets can resonantly capture neutrinos using beam energies not significantly higher than those currently attainable at the LHC. Moreover, the neutrino capture cross sections for these resonant processes scale quadratically with the inverse of the energy threshold of the process, favouring experiments at accelerators with lower beam energies.
 The resonance states are unstable, but can be observed through decays into a stable final state different from the initial state. We show how to find isobar triplets with the required properties to be 3-state system candidates. We have calculated the expected population of resonance and final states on the beam for both 2 and 3-state systems as a function of the experimental runtime and provide explicit examples. 
 
Even though the experiment can be performed in principle, significant experimental challenges remain. The biggest challenge is to find an ion with threshold as small as possible, $Q\lesssim 10\,\mathrm{keV}$. Because of the associated synchroton radiation, it is difficult to achieve the large number of ions on the beam required to achieve reasonable statistics, which far exceeds the number on any existing accelerator experiment. Systems with lower energy thresholds are again favoured, as the total power emitted by synchroton radiation scales like the fourth power of the energy threshold. In contrast to a collider experiment, the ions need not collide, which may allow for a modified design with more ions on the beam.

As precisely hitting the resonance requires knowledge of the neutrino mass, a future measurement with low uncertainty would allow us to considerably refine the predictions made in this work. Further exploration of systems capable of resonantly capturing neutrinos may also yield promising results. 

We have not identified a system capable of discovering the C$\nu$B with current technology. However, the systems proposed for resonant neutrino capture here are not an extensive list and it is highly likely that a much better system can be found.
Ongoing experiments like KATRIN will provide further input within the next 5 years which could make resonant neutrino capture a promising method of discovering the cosmic neutrino background. 

The resonant neutrino capture processes we discussed could be utilised with exotic systems or an entirely different experimental setup. Additionally, this technology has the potential to detect other sources of low energy neutrinos, such as those from thermal solar processes or new physics such as majoron decays~\cite{Chacko:2018uke}.

\section{Acknowledgements}
The authors would like to thank Edoardo Vitagliano for helpful discussions regarding the C$\nu$B neutrino fluxes. We would also like to thank Yeongduk Kim for bringing the updated electron binding energies to our attention. Martin Bauer thanks Yuval Grossmann for encouraging discussions on this topic. We further thank Stewart Boogert for inventing the ``death beam limit". Jack D. Shergold is supported by an STFC studentship under the STFC training grant ST/T506047/1.

\appendix

\section{Bound beta decay rates}\label{App:bbdecay}
\begin{figure}[t]
    \centering
    \includegraphics[width=\linewidth]{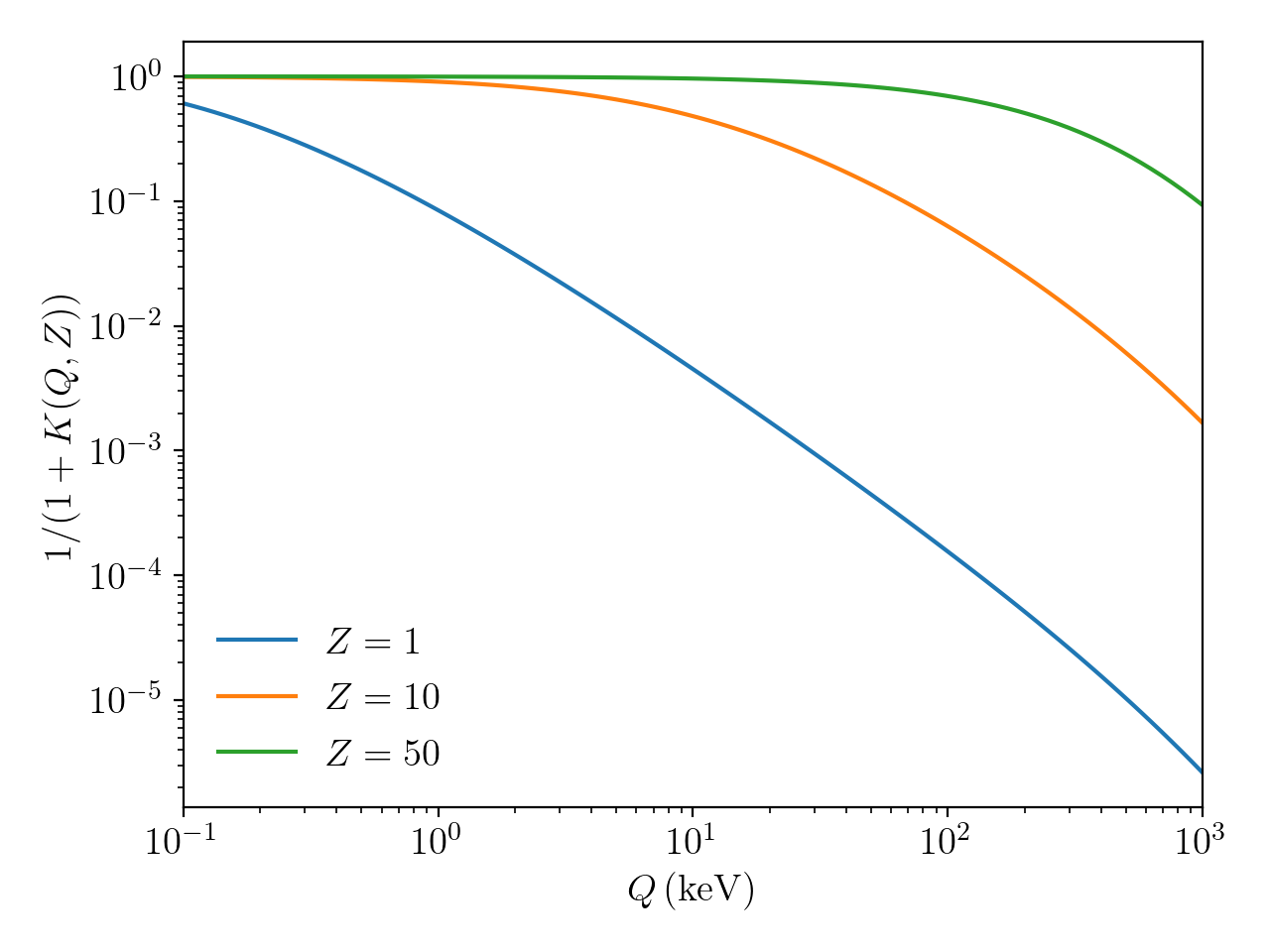}
    \caption{Dependence of the bound beta decay branching ratio on $K(Q,Z)$. We choose $A = 2Z, n_f = 2$ and set $Q_{\mathrm{C}\beta} = Q_{\mathrm{B}\beta} = Q$.}
    \label{fig:kFunc}
\end{figure}

For the ratio of continuous to bound beta decay, we use the expression in~\cite{Oldeman:2009wa} for an electron captured in a shell with principal quantum number $n$ with $n_f$ free orbitals,
\begin{equation}
    \frac{\Gamma_{\mathrm{C}\beta}}{\Gamma_{\mathrm{B}\beta}} = \left(\frac{n}{\alpha}\right)^3 \frac{1}{n_f \pi} \frac{f(Z,Q_{\mathrm{C}\beta})}{Z^{2.87 + (6.2\times10^{-3})Z}}\left(\frac{m_e}{Q_{\mathrm{B}\beta}}\right)^2,
\end{equation}
where $Q_\mathrm{C\beta}$ is the $Q$-value for continuous beta decay and we define the function
\begin{align}
        f(Z,Q) = \frac{1}{m_e^5}\int\displaylimits_0^Q & d\widetilde{E}_{k,e} (Q-\widetilde{E}_{k,e})^2 (\widetilde{E}_{k,e} + m_e)\\ 
        &\times \sqrt{\widetilde{E}_{k,e}^2 + m_e \widetilde{E}_{k,e}} F_{-}(Z, \widetilde{E}_{k,e} + m_e).\notag
\end{align}
As we only consider beta-decaying states that are either fully ionised or have one electron, the branching ratios in Section~\ref{sec:resonant} depend on the ratio 
\begin{equation}
    K(Z,Q) = \frac{\Gamma_{\mathrm{C}\beta}}{\Gamma_{\mathrm{B}\beta}}\bigg|_{n = 1}.
\end{equation}
We show the dependence of the bound beta decay branching ratios on $K(Q,Z)$ in Figure~\ref{fig:kFunc}, which displays a strong dependence on the $Q$-value and the atomic number $Z$.

\begin{table}
\renewcommand{\arraystretch}{1.5}
    \begin{tabular}{c|c|c|c}
        Source & $E_{k,\text{max}}\,$(keV) & $\phi_\mathrm{LZ}\,$(\SI{}{\per\centi\metre\squared\per\second}) & $\phi_\mathrm{HZ} \,$(\SI{}{\per\centi\metre\squared\per\second}) \\ \hline \hline
        $pp$ & 420 & $6.03\times10^{10}$ & $5.98\times10^{10}$ \\
        $^{7}$Be & 384 (es), 862 (gs) & $4.50\times10^{9}$ & $4.93\times10^{9}$ \\
        $^{13}$N & 1198 & $2.04\times10^{8}$ & $2.78\times10^{8}$ \\
        $^{15}$O & 1732 & $1.44\times10^{8}$ & $2.05\times10^{8}$ \\
        $pep$ & 1440 & $1.46\times10^{8}$ & $1.44\times10^{8}$ \\
        $^{8}$B & $1.70\times10^{4}$ & $4.50\times10^{6}$ & $5.46\times10^{6}$ \\
        $^{17}$F & 1738 & $3.26\times10^{6}$ & $5.29\times10^{6}$ \\
        He$p$ & $1.88\times10^{4}$ & $8.25\times10^{3}$ & $7.98\times10^{3}$
    \end{tabular}
    \caption{Non-thermal solar neutrino fluxes at Earth. The abbreviations gs and es refer to the decays of $^7$Be to the ground state and excited state of $^7$Li respectively.}
    \label{tab:solarFlux}
\end{table}

\section{Generation of the solar neutrino flux}\label{App:solarneutrinos}

For our calculations of the solar neutrino fluxes given in Figure \ref{fig:fluxes}, we assume the solar chemical composition of ~\cite{Asplund}, which we refer to as the low-metallicity (LZ) standard solar model herein. The LZ model is able to accurately reproduce spectroscopic data, however, it falls into tension with helioseismology ~\cite{Vinyoles:2016djt, Agostini:2018uly}. On the contrary, the model given in~\cite{Grevesse:1998bj} uses a higher solar metallicity (HZ) and is in better agreement with helioseismology. As such, we show the predicted neutrino fluxes from both the LZ and HZ models in Table \ref{tab:solarFlux} for completeness. Importantly, we note that as the predictions are all of the same order of magnitude, the conclusions of this work are unaffected by the choice of model. 

Considering the three body decay of an unstable ion in the sun (e.g $^{15}\mathrm{O}\rightarrow ^{15}\mathrm{N} + e^+ + \nu_e$) where the final state nuclear recoil and small neutrino mass are neglected, the associated electron neutrino flux in the lab frame is

\begin{equation}
    \begin{split}
        \frac{d\phi}{dE_{k,\nu}} \simeq \mathcal{N}\, C&(E_e) F_+(Z,E_e) E_{k,\nu}^2 (Q-E_{k,\nu} + m_e) \\
        & \times \sqrt{(Q-E_{k,\nu} + m_e)^2 - m_e^2},
    \end{split}
\end{equation}
where the positron energy is given by $E_e \simeq m_e + Q - E_{k,\nu}$, $C(E_e)$ is the nuclear shape factor and $\mathcal{N}$ is a normalisation factor, calculated numerically such that the integrated fluxes agree with those in Table \ref{tab:solarFlux}. For allowed and superallowed transitions the shape factor is approximately constant with energy, and so can be safely absorbed into the normalisation factor~\cite{Cocco:2007za, Cocco:2008nka, Behrens:1971rbq}.

\begin{figure}[t]
    \centering
 \includegraphics[width = 1.\linewidth]{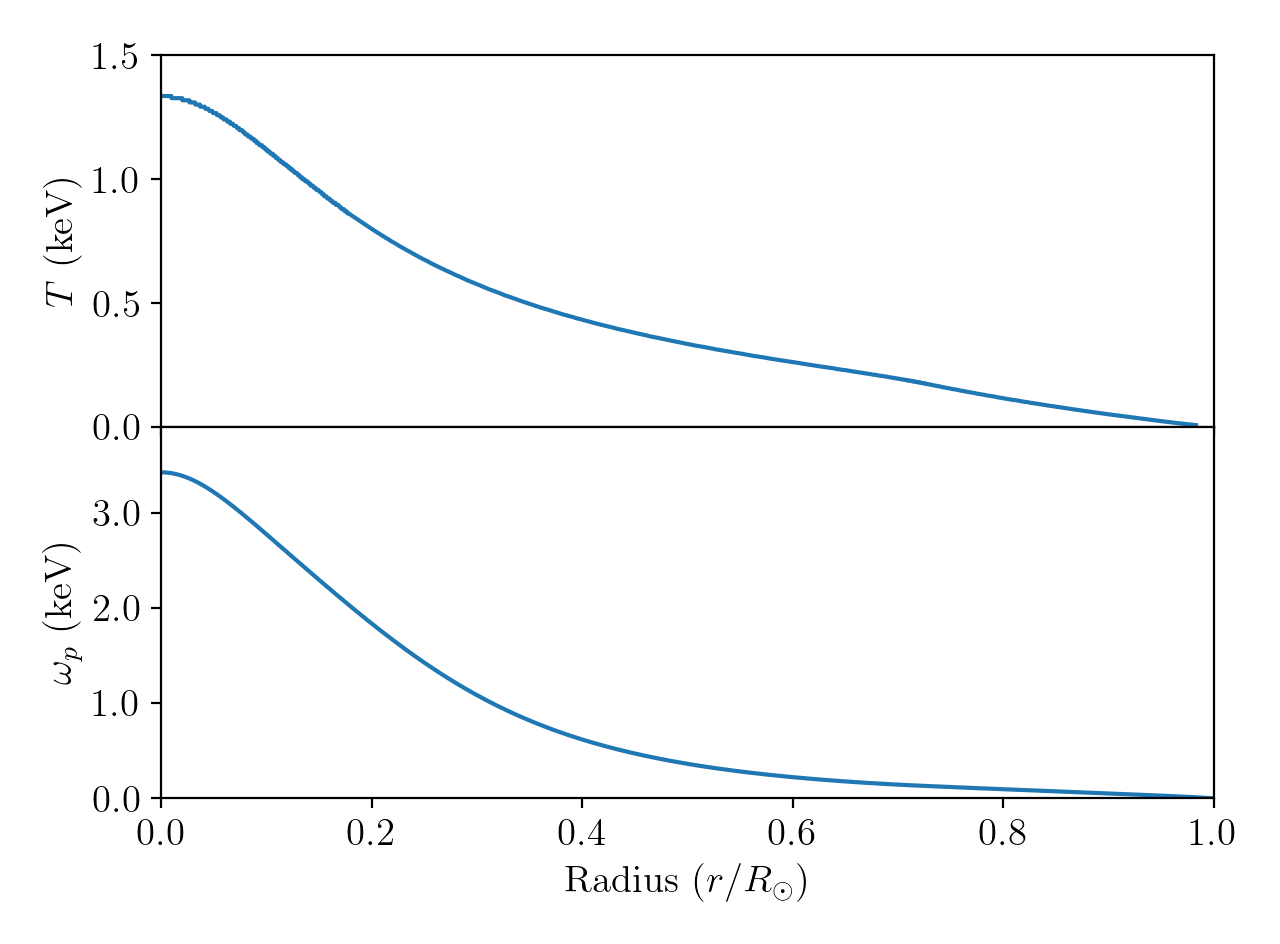}
 \caption{Temperature and plasma frequency profiles of the sun.}
    \label{fig:sunProfiles}
\end{figure}

In addition to those neutrinos generated by decays in the sun, we also consider the flux of thermal solar neutrinos, which dominate the spectrum in the \SI{}{\electronvolt} energy range. The largest such contribution comes from plasmon decays $\gamma \rightarrow \bar{\nu} \nu$, for which the flux differs depending on the polarisation of the plasmon. A full treatment of thermal sources of neutrinos should also consider both the bremsstrahlung $e\rightarrow e\bar{\nu}\nu$ and photo-production $e\gamma\rightarrow e\bar{\nu}\nu$ processes~\cite{Vitagliano:2017odj}, however, as the neutrino fluxes from these sources are mostly subdominant to the plasmon decay flux we do not consider them here.

Following the formalism of~\cite{Vitagliano:2017odj} and introducing the dimensionless variable $\ell = r/R_{\odot}$ in terms of the distance from the centre of the sun $r$, the flux of electron neutrinos from transverse (T) plasmon decay reaching Earth is
\begin{align}
    \frac{d\phi_{\nu_e}}{dE_{\nu}}\bigg|_\text{T}&=\frac{\sum_{i}|U_{ei}|^4}{64\pi^3 \alpha}\frac{R_{\odot}^3 C_V^2 G_F^2}{(1\text{AU})^2}\\
    &\times\int\displaylimits_0^1 d\ell\int\displaylimits_{\omega_0(x)}^\infty d\omega \, \frac{\ell^2 \omega_p(\ell)^6}{e^{\omega/T(\ell)}-1}\bigg[1+\frac{(\omega-2E_{\nu})^2}{\omega^2 - \omega_p(\ell)^2}\bigg],\notag
\end{align}
where the 1AU denotes one astronomical unit, $R_\odot$ is the solar radius, and $T(\ell)$ is the solar temperature. The lower bound of the integral can be written in terms of the plasma frequency $\omega_p(\ell)$ as
\begin{equation}
    \omega_0(\ell) = E_{\nu}+\frac{\omega_p^2(\ell)}{4E_{\nu}}.
\end{equation}
Similarly, one finds for longitudinally polarised (L) plasmons
\begin{equation}
    \frac{d\phi_{\nu_e}}{dE_{\nu}}\bigg|_\text{L}=\frac{\sum_{i}|U_{ei}|^4}{64\pi^3 \alpha}\frac{R_{\odot}^3 C_V^2 G_F^2}{(1\text{AU})^2} \int\displaylimits_0^1 d\ell \, \frac{x^2 \omega_p(\ell)^7P(\ell)}{e^{\omega_p(\ell)/T(\ell)}-1} ,
\end{equation}
where 
\begin{align}
\hspace{-.5cm}P(\ell)= \frac{2+3y(\ell)^2-6y(\ell)^4+y(\ell)^6}{12}+y(\ell)^2\log|y(\ell)|\,,
\end{align}
 $y(\ell) = (2E_{\nu}-\omega_p(\ell))/\omega_p(\ell)$, and the condition $0<E_\nu<\omega_p(\ell)$ is also enforced. In both cases, $C_V^2 \simeq 0.9263$ is the effective vector coupling between electrons and electron neutrinos, and we have neglected the small contribution from the heavy lepton neutrino flavours. The temperature and plasma frequency profiles used are shown in Figure~\ref{fig:sunProfiles} and were calculated using the BS05(AGS,OP) solar model outlined in~\cite{Bahcall:2004pz}, for which the
elemental abundances are similar to those of the LZ model. 

\section{Number of ions on a beam}\label{App:beamstates}

To find the number of each ion state on the beam at a given time, we need to take into account both the effects of neutrino captures and the decays of unstable states, which act to populate and empty states respectively. We neglect the universal decay of the beam due to continuous beam losses. 

We calculate the beam populations for a 3-state system with an excited initial state, from which all other systems can be derived by taking the appropriate limits. The number of $P$ states on the beam at any one time can be found by solving the differential equation 
\begin{equation} \label{eq:NP2stateDE}
    \frac{dN_P}{d\tilde{t}} =-\left(\gamma \frac{R}{N_T} + \frac{1}{\tau_{P^*}}\right) N_P(\tilde{t})+ \frac{\mathcal{B}_{DP}}{\tau_D} N_D(\tilde{t}),
\end{equation}
in terms of the beam rest frame time $\tilde t = t/\gamma$. The factor of $\gamma$ appearing before the term proportional to $R$ is due to the increased neutrino flux in the beam rest frame. 
This can be re-expressed in terms of dimensionless parameters, the quality factor $R_\tau$, the parent-daughter lifetime ratio $\eta$ and $x$ as
\begin{equation}
    \frac{dN_P}{dx}= - \left(R_\tau + \eta\right) N_P(x) + \mathcal{B}_{DP} N_D(x)\,,
\end{equation}
where we have used $\Gamma = 1/\tau_D$. The analogous differential equations for the number of $D$ and $F$ states on the beam in terms of the same dimensionless parameters are given by
\begin{align}
    \frac{dN_D}{dx} &= R_\tau N_P(x) - N_D(x), \\
    \frac{dN_F}{dx} &= \mathcal{B}_{DF} N_D(x).
\end{align}
Using the initial conditions $N_P(0) = N_0$, $N_D(0) = 0$ and $N_F(0) = 0$ yields the number of $D$ and $F$ states on the beam at a given value of $x$,
\begin{align}
    N_D(x) = &\frac{2 N_0 R_\tau}{\kappa}\sinh\left(\frac{\kappa x}{2}\right) e^{-\frac{x}{2}(1+R_\tau + \eta)}\, , \label{eq:xySignal}\\
    N_F(x) = &\frac{N_0 \mathcal{B}_{DF} R_\tau}{\eta + R_\tau (1-\mathcal{B}_{DP})}\bigg[1-e^{-\frac{x}{2}(1+R_\tau + \eta)} \label{eq:NF3states}\\ 
    \times & \left\{\left(\frac{1 + R_\tau + \eta}{\kappa}\right)\sinh\left(\frac{\kappa x}{2}\right) + \cosh\left(\frac{\kappa x}{2}\right)\right\}\bigg]\,,\notag 
\end{align}
where in both cases $\kappa = \sqrt{(1-R_\tau - \eta)^2 + 4 \mathcal{B}_{DP} R_\tau}$. To recover the 2-state system equations, one simply sets $\mathcal{B}_{DF} = 0$. The ground state systems can be obtained from \eqref{eq:xySignal} and \eqref{eq:NF3states} by taking the limit $\eta \to 0$ and replacing $\mathcal{B}_{DP}$ with $\mathcal{B}_{DP}/\chi$. For ground state systems we divide the branching ratio $\mathcal{B}_{DP}$ by the probability to decay into the initial state isomer $\chi$ as excited state isomers produced by resonance decays on the beam will quickly decay to the ground state isomer, becoming available to capture neutrinos again.


\end{document}